\documentstyle[preprint,prb,aps,epsf,eqsecnum]{revtex}
\begin{document}
\draft

\title{Dynamic correlations of antiferromagnetic spin-1/2 XXZ chains
at arbitrary temperature from complete diagonalization }

\author{Klaus Fabricius}
\address{Physics Department, University of Wuppertal, 42097 Wuppertal,
Germany} 
\author{Ute L\"ow}  
\address{Physikalisches Institut der Johann Wolfgang Goethe Universit\"at,
D-60054 Frankfurt am Main, Germany}
\author{Joachim Stolze}
\address{Institut f\"ur Physik, Universit\"at Dortmund, 44221
Dortmund, Germany}

\maketitle

\begin{abstract}
All eigenstates and eigenvalues are determined for the spin- 1/2 $XXZ$
chain $H = 2J \sum_i ( S_{i}^{x} S_{i + 1}^{x} + S_{i}^{y} S_{i +
1}^{y} + \Delta S_i^z S_{i + 1}^{z})$ for rings with up to $N=16$
spins, for anisotropies  $\Delta=0 , \cos(0.3
\pi)$, and $1$.  The dynamic spin pair
correlations $\langle S_{l+n}^{\mu}(t) S_l^{\mu} \rangle , (\mu=x,z)$,
the dynamic structure factors $S^{\mu}(q,\omega)$, and the
intermediate structure factors $I^{\mu}(q,t)$ are calculated for
arbitrary temperature $T$. It is found, that for all $T$, 
$S^{z}(q,\omega)$ is mainly
concentrated on the region $|\omega| < \varepsilon_2(q)$, where
$\varepsilon_2(q)$ is the upper boundary of the two-spinon continuum,
although excited states corresponding to a much broader frequency
spectrum contribute.  This is also true for 
the Haldane-Shastry model and the frustrated Heisenberg model. 
The intermediate structure factors $I^{\mu}(q,t)$ for $\Delta
\neq 0$ show exponential decay for high $T$ and large $q$.
Within the accessible time range, the time-dependent spin
correlation functions do not display the long-time signatures of spin
diffusion.
\end{abstract}



\vskip 0.3 truein

\section{Introduction}
\label{I}

We discuss the one-dimensional $S=1/2$ $XXZ$ antiferromagnet,
specified by the Hamiltonian 
\begin{equation}
H_{XXZ} = 2 J \sum_{i = 1}^{N} \left\{ S_{i}^{x} S_{i + 1}^{x} +
S_{i}^{y} S_{i + 1}^{y} + \Delta S_i^z S_{i + 1}^{z} \right\} \, 
\label{I.1},
\end{equation}
with coupling $J > 0$ and anisotropy $0 \leq \Delta \leq 1$. The model is
defined on a ring of $N$ lattice sites. 
The quantities studied in this paper
 are the dynamic spin pair correlation functions
$\langle S_l^{\mu} (t)S_{l+n}^{\mu} \rangle \quad (\mu=x,z)$ and the
quantities which can be obtained from 
$\langle S_l^{\mu} (t)S_{l+n}^{\mu} \rangle$ 
by partial or complete Fourier
transformation with respect to space and time, namely the spatial
Fourier transform, which is sometimes also called intermediate
structure factor,
\begin{equation}
I^{\mu}(q,t) =   \sum_n \exp(-iqn) 
\langle S_l^{\mu} (t)S_{l+n}^{\mu} \rangle
\label{I.2},
\end{equation}
the temporal Fourier transform
\begin{equation}
\Phi^{\mu}_n(\omega) =  \int_{-\infty}^{\infty} dt \exp(i \omega t)
\langle S_l^{\mu} (t)S_{l+n}^{\mu} \rangle
\label{I.3},
\end{equation}
and finally the dynamic structure factor
\begin{equation}
S^{\mu}(q,\omega) =   \sum_n 
\int_{-\infty}^{\infty} dt \exp(-i (q n - \omega t))
\langle S_l^{\mu} (t)S_{l+n}^{\mu} \rangle
\label{I.4},
\end{equation}
which is related to the inelastic neutron scattering cross section
\cite{BL89}. We have determined the above quantities by complete
diagonalization of the Hamiltonian (\ref{I.1}) for system sizes $N
\leq 16$ and anisotropies $\Delta = 0, \cos 0.3 \pi$, and 1, at
various temperatures.
Given the eigenvectors $|\nu \rangle$ of $H_{XXZ}$ and the
corresponding eigenvalues $E_{\nu}$, the dynamic structure factor may
be computed as
\begin{equation}
S^{\mu}(q,\omega) =   \frac{2 \pi}{Z} \sum_{\nu, \lambda}
\exp(- \beta E_{\nu}) \delta(\omega - E_{\lambda} + E_{\nu})
|\langle \nu| S_q^{\mu} |\lambda\rangle|^2
\label{I.5},
\end{equation}
where
\begin{equation}
Z=\sum_{\nu}\exp(- \beta E_{\nu})
\label{I.6}
\end{equation}
is the partition function, and 
\begin{equation}
S_q^{\mu} = \frac{1}{\sqrt N} \sum_{l=1}^N \exp(-i q l) S_l^{\mu}
\label{I.7}.
\end{equation}

The plan of the paper is as follows. In Sec. \ref{II} we give a survey of
previous (analytic) results, for the $XX$ model  (\ref{IIA}) and
for the $XXZ$ model (\ref{IIB}), we also recall the phenomenological
pictures of ballistic and diffusive dynamics (\ref{IIC}). 
Sec. \ref{VA} addresses the question, which matrix elements give
the largest contributions to the dynamic structure factor as 
defined in Eq. (\ref{I.5}). We have studied this question 
in some detail for the $XXX$-model ($\Delta=1$) and, for purposes of
comparison, also for the frustrated Heisenberg model and
the Haldane-Shastry model. 
In Sec. \ref{VB} we present our results for the dynamic structure 
factors. Sec. \ref{VD} contains remarks on finite-size effects.
Sec. \ref{III} is devoted to
time-dependent correlation functions
and Sec. \ref{VI} contains a summary.
A detailed comparison of our
numerical results (including also effects of a nonzero magnetic field) 
to recent experimental data from spin-chain
compounds like KCuF$_3$ or CuGeO$_3$ is planned for a later publication.

\section{Previous results}
\label{II}

\subsection{The $XX$ model}
\label{IIA} 

For the $XX$ model ($\Delta=0$), which can be mapped to a system of
noninteracting lattice fermions \cite{LSM61,K62} by the Jordan-Wigner
transformation, time-dependent correlation functions have been
calculated
\cite{N67,KHS70,MBA71,VT78,MPS83,MS84,SJL75,BJ76,CP77,IIKS93}
 at zero and nonzero
temperatures. In that case the spin correlation function $\langle
S_i^z (t)S_j^z \rangle$ is a simple fermion density correlation
function, and the function $\langle S_i^x (t)S_j^x \rangle$ can be
reduced to a determinant whose size increases linearly with $i+j$.
 
Niemeijer \cite{N67} and Katsura et al. \cite{KHS70} derived 
explicit expressions for the correlation
$\langle S_l^z (t)S_{l+n}^z \rangle$ of the $XX$ model and the 
associated quantities (\ref{I.2}-\ref{I.4}) which we will recall 
below.

For arbitrary $T$ and infinite N
the
structure factor is
\begin{eqnarray}
\lefteqn{S^z(q,\omega)  =   \frac
      {\Theta(|4J \sin(\frac{q}{2})| - |\omega|)}
      {2 [(4 J)^2 \sin^2 (\frac{q}{2}) - \omega^2]^{\frac{1}{2}}}}
\nonumber \\
 & &    \frac{\exp(\frac{\beta \omega}{2})}
        { \cosh \left[ \frac{\beta \omega}{4} +
        \beta  J \cos(\frac{q}{2}) [1 - \frac{\omega^2}{(4 J)^2 \sin^2
        (\frac{q}{2})}]^{1/2}\right]} \nonumber \\
 & &
\frac{1}
        { \cosh \left[ \frac{\beta \omega}{4} -
        \beta  J \cos(\frac{q}{2}) [1 - \frac{\omega^2}{(4 J)^2 \sin^2
        (\frac{q}{2})}]^{1/2}\right]}
      \label{II.6}.
\end{eqnarray}
At the upper boundary of the frequency spectrum 
$S^z(q,\omega)$ displays an inverse square-root
divergence for all temperatures. This divergence translates into an
asymptotic $t^{-1/2} \exp(i 4 J t \sin(q/2))$ behavior for $I^z(q,t)$ and
the existence of a high-frequency cutoff also leads to the long-
time asymptotic behavior $\langle S_l^z(t) S_{l+n}^z \rangle \sim 
t^{-1}$. At $T=0$, the dynamic structure factor $S^z(q,\omega)$ as given in 
(\ref{II.6}) has also a lower spectral boundary, where it displays 
a discontinuity. 
The longitudinal spin pair correlations at infinite $T$ are
\begin{equation}
\langle S_l^z(t) S_{l+n}^z \rangle = \frac{1}{4} J_n^2(2Jt)
      \label{II.1}.
\end{equation}
and \cite{GR80}
\begin{equation}
I^z(q,t)=\frac{1}{4} J_0(4Jt \sin\frac{q}{2})
      \label{II.2}.
\end{equation}

The transverse correlation 
functions $\langle S_l^x(t) S_{l+n}^x\rangle$ of the $XX$ chain 
are complicated many-particle correlations of the Jordan-Wigner 
fermions and fewer explicit results are available than 
for the longitudinal
correlations. In fact, the only simple closed-form result is
\cite{SJL75,BJ76,CP77} 
\begin{equation}
S^x(q,\omega) =  \frac{\pi^\frac{1}{2}}{4J} \exp(-\frac{\omega^2}{4J^2})
\label{II.12},
\end{equation}
for $ T=\infty $. 
Its et al. \cite{IIKS93} determined $\langle S_l^x(t) S_{l+n}^x\rangle$
for $t\rightarrow\infty$, $n\rightarrow\infty$ and $n/t=\mbox{const}$ in a 
moderate magnetic field.

\subsection{The $XXZ$ model}
\label{IIB}

At $T=0$ the long-distance and long-time properties of the $XXZ$
chain may be described by a Fermi field theory \cite{LP75,F78}.
This leads to power-law singularities with $\Delta$-dependent
exponents for several quantities. Specifically we have
\begin{equation}
\langle S_l^{\mu}(t) S_{l+n}^{\mu} \rangle \sim (-1)^n [n^2 -v^2
t^2]^{-\frac{1}{2}\eta_{\mu}} 
\label{II.10}
\end{equation}
with
\begin{equation}
\eta_x = \eta_z^{-1} = 1 - \frac{\gamma}{\pi}
\label{II.11},
\end{equation}
where
\begin{equation}
\Delta = \cos \gamma
\label{II.9},
\end{equation}
and 
\begin{equation}
v=\frac{\pi J \sin \gamma}{\gamma}
\label{II.9a}.
\end{equation}
This leads\cite{18neu} to 
corresponding low-frequency singularities in $S^{\mu}(q,\omega)$ 
and $\Phi^{\mu}_0(\omega)$. 
The structure of these singularities
has been the subject of a recent study \cite{FML95}.

The continuum of $(q,\omega)$ values for which $S^z(q,\omega,T=0)$
is nonzero in the $XX$ model is a special case of a more
general continuum in the $XXZ$ case.
The spectral boundaries of this  
two-spinon continuum are given by

\begin{mathletters}
\begin{equation}
\varepsilon_1(q) = \frac{\pi J \sin \gamma}{ \gamma} \sin q
\label{II.8a},
\end{equation}
\begin{equation}
\varepsilon_2(q) = \frac{2 \pi J \sin \gamma}{ \gamma} \sin \frac{q}{2}
\label{II.8b}.
\end{equation}
\end{mathletters}
The nature of the two-spinon states defining the continuum will be
discussed in Sec. \ref{VA}.
It was argued \cite{MTPB81}
 that at $T=0$ the dominant contributions
to both $S^x(q,\omega)$ and $S^z(q,\omega)$ come from this 
continuum and a related one with the same lower boundary and the 
upper boundary 
\begin{equation}
\tilde{\varepsilon}_2(q) = \varepsilon_2(\pi -
 q) \label{tilde}.
\end{equation}
Approximate analytic expressions for the $T=0$ dynamic structure
factors were conjectured, which take into account known sum rules as
well as exact results for the case $\Delta=0$ and asymptotic results
(\ref{II.10}) for small $q$ and $\omega$.  
For explicit formulae and further discussion we refer the reader to
Ref.\onlinecite{MTPB81}. Very recently \cite{BCK96,KMB96} the exact
two-spinon contribution to $S^{\mu} (q,\omega)$ of the Heisenberg
model has been  be calculated analytically at $T=0$ and it was found that
it accounts for more than 80 percent of the total intensity.

Schulz \cite{S86} used field-theoretical techniques to
study a spin-$S$ $XXZ$ antiferromagnetic chain at finite (not too
high) $T$. He derived approximate analytic expressions for
$S^{\mu}(q,\omega)$ valid for small values of $\omega$ and $|q-\pi|$,
which showed good agreement to recent $T$-dependent inelastic neutron
scattering data \cite{TCNT95} from the $S=1/2$ Heisenberg
antiferromagnet KCuF$_3$.
For the case of interest, the results for $S^{\mu}(q,\omega)$
derived by Schulz 
may be written as 
\begin{eqnarray}
\lefteqn
{
S^{\mu}(q,\omega) = A_{\mu}(\gamma) (n_{\omega} + 1)
\mbox{Im} \left\{ 
 \sin(\frac{\pi\eta_{\mu}}{2})
\Gamma^2(1-\frac{\eta_{\mu}}{2}) 
 \right. 
}
\nonumber\\
 & &  \left. \left[ \frac{2 \pi \alpha T}{v} \right] ^{\eta_{\mu}-2}
\rho(x_+) \rho(x_-) - \frac{\pi \delta_{\mu z}}{1 -
\frac{\eta_{\mu}}{2}}
\right\}
\label{S1}.
\end{eqnarray}
Here 
\begin{equation}
 n_{\omega} = (\exp(\beta \omega) - 1)^{-1},\ \ \ 
\rho(x)=\frac{\Gamma(\frac{\eta_{\mu}}{4} -ix)} 
               {\Gamma(1-\frac{\eta_{\mu}}{4} - ix)} , \ \ \ 
x_{\pm} = \frac{\omega \pm v (q-\pi) }{4 \pi T}
\end{equation}
and
$v$ and $\eta_{\mu}$ are defined in (\ref{II.9a}) and
(\ref{II.11}). 
$\alpha$ is a short-distance cutoff (comparable to the lattice
constant), and $A_{\mu}(\gamma)$ is an overall factor.  
The $\delta_{\mu z}$ term serves to cancel a spurious divergence
which occurs for $\mu=z$ as $\gamma \rightarrow
\frac{\pi}{2}$. 
At $T=0$, (\ref{S1}) yields
\begin{equation}
S^{\mu}(q,\omega) \sim \Theta(\omega- |v \tilde{q}|) (\omega^2- (v
\tilde{q})^2)^{\frac{\eta_{\mu}}{2} - 1}
\label{S3}.
\end{equation}
This expression correctly reproduces the $T=0$ singularities
\cite{LP75,F78,MTPB81} at the lower continuum boundary
$\varepsilon_1(q)$ (\ref{II.8a}). Note, however, that the
experimentally confirmed \cite{TCNT95} upper continuum boundary
$\varepsilon_2(q)$ (\ref{II.8b}) does not show up in
(\ref{S3}). Consequently the frequency integral over
$S^{\mu}(q,\omega)$ diverges for arbitrary $q$ and $\gamma$. 

Schulz' field theoretical approximation (\ref{S1}) reads in 
case of the $XX$-model and $q=\pi$ 
\begin{equation}
S^z(q=\pi,\omega, T) \propto 
      \frac
      {1}
      {(1 + \exp( \frac{-\beta \omega}{2}))^2}
      \label{II.3}
\end{equation}
which is the exact result Eq.(\ref{II.6}) up to the factor 
\begin{equation}
\frac
      {\Theta(4J - |\omega|)}
      {[(4 J)^2  - \omega^2]^{\frac{1}{2}}}
      \label{II.3a}.
\end{equation}

That is, Schulz' formula describes the $T$ dependence of
the exact $XX$ result correctly, but fails to reproduce the cutoff and
 the square root singularity in $\omega$ at the cutoff.

\subsection{Phenomenological pictures: Ballistic and diffusive behavior}
\label{IIC}

Spin diffusion has been a popular \cite{SVW76} 
concept used in the discussion of experimental results 
for decades.
According to the spin diffusion hypothesis one expects
\begin{equation}
\langle S_l^{\mu}(t) S_{l+n}^{\mu} \rangle - \langle S_l^{\mu}
      S_{l+n}^{\mu} \rangle \sim t^{-\frac{1}{2}} \exp(-
      \frac{n^2}{4Dt})
\label{II.16}
\end{equation}
for long times and long distances, where $D$ is the diffusion constant.
The intermediate scattering function correspondingly is 
\begin{equation}
I^{\mu}(q,t) - I^{\mu}(q,0) \sim \exp(-Dq^2t)
\label{II.17}
\end{equation}
for small $q$ and long times. 

A quantity sensitive to spin diffusion, the ``spatial variance'', was
introduced by B\"ohm and Leschke \cite{BL92}. It is defined as follows:
\begin{equation}
\sigma_{\mu}^2(t)=4\sum_n ( \langle S_l^{\mu}(t) S_{l+n}^{\mu} \rangle
     - \langle S_l^{\mu} S_{l+n}^{\mu} \rangle ) n^2
\label{II.19}.
\end{equation}
Specifically, $\sigma_{\mu}^2(t)$ allows to distinguish 
between diffusive and ballistic behavior of the spin correlations.
In the diffusive case, the intermediate structure factor
(\ref{II.17}) yields
\begin{equation}
- \left. \frac{\partial^2}{\partial q^2}  (I^{\mu}(q,t) - I^{\mu}(q,0))
\right|_{q=0}
= \frac{1}{4} \sigma_{\mu}^2(t) \sim 2Dt
\label{II.20}
\end{equation}
On the other hand, ballistic propagation leads to
\begin{equation}
\sigma_{\mu}^2(t) \sim t^2
\label{II.21}
\end{equation}
with the mean square velocity as a factor of proportionality.

For the $XX$ chain and $T=\infty$ one gets 
\begin{equation}
\sigma_x^2(t)=0
\label{II.22}
\end{equation}
and from (\ref{II.2}) one obtains, by  (\ref{II.20})
\begin{equation}
\sigma_z^2(t) = 2 J^2 t^2
\label{II.23},
\end{equation}
meaning that in that special case the conserved $z$ spin component
propagates ballistically, whereas the non-conserved $x$ component 
decays locally. 
At finite $T$ one obtains
\begin{equation}
\mbox{Re} \sigma_z^2(t) \propto t^2 \quad,\quad \mbox{Im} 
 \sigma_z^2(t) \propto t
\label{II.24}
\end{equation}
for $\Delta=0$ and arbitrary $t$. It should be noted that for
{\em short} times, (\ref{II.24}) is the leading-order behavior to be
expected for both $\sigma_x^2(t)$ and  $\sigma_z^2(t)$ (and for zero
as well as nonzero $\Delta$) due to the general
symmetry $\langle S_l^{\mu}(t) S_{l+n}^{\mu} \rangle = 
\langle S_l^{\mu}(-t) S_{l+n}^{\mu} \rangle^* $. From the asymptotic
behavior found by Its et al. \cite{IIKS93}
$\sigma_x^2(t)$ is expected to show
exponentially damped oscillatory behavior for long times.

Numerous theoretical studies have attacked the question whether spin
diffusion describes the high-temperature dynamics of quantum spin
chains; we will quote only a few selected references. 
In an
important early numerical study, Sur and Lowe \cite{SL75} computed the
$XXX$ spin autocorrelation function at $T=\infty$ for $N\leq11$. A
comprehensive $T=\infty$ moment study \cite{RMP86} was extended a few
years later \cite{BL92}.  It was conjectured\cite{M95}  that 
spin diffusion is absent in the $XXZ$ chain
and numerical evidence to that effect was reported
recently\cite{ZP96}.
Other recent publications employ different approximate methods and
arrive at different conclusions. In Ref.\onlinecite{N96} spin diffusion
is not found, whereas in Refs.\onlinecite{BVSM94,BLHVSM94} 
spin diffusion is found.
We have studied the longest chains to date, and for the time
range covered in our study, the $T=\infty$ results presented in
Sec. \ref{III} do not show the signatures of spin diffusion.

\section{Dynamic structure factor}
\label{V}

\subsection{Excitation continua and classes of states}
\label{VA}

Here we shall first remind the reader of the main 
conclusions at $T=0$  and then describe the corresponding results for $T>0$.

For the $T=0$ spin dynamics a very simple picture has emerged : 
only transitions which excite not more than two spinons are of importance.
This statement will be made more precise below.
In the following we have to use the Bethe-Ansatz \cite{Beth31}
terminology (see appendix).
Excitations with a small number $m$ of flipped spins with respect to
the ferromagnetic ground state have spin wave character and are called 
magnon excitations. 
They are relevant for the ferromagnetic model and for very strong
magnetic fields in the antiferromagnetic model.
For excitations of the antiferromagnetic ground 
state with $m \approx N/2 $ and large $N$  
a description in terms of  magnons is inappropriate.
These states are fully characterized by the positions of holes in 
their $ \lambda$ - sequence. Excitations of this kind have
first been studied by des Cloizeaux and Pearson \cite{ClPe62}. 
The first complete
determination of the two parameter continuum 
corresponding to two independent holes in the $\lambda$ sequence
of $S=1$ states was 
presented by Yamada \cite{Yam69} and M\"uller et al. \cite{MTBB}. 
A systematic description
was later given by  Faddeev and Takhtajan \cite{FaTa81,FaTa84}. 
Some important examples were 
worked out for the anisotropic model by Woynarovich \cite{Woy82}. 
These authors found that the elementary
excitations of the antiferromagnetic Heisenberg model have spin 1/2
and that their number is always even. In recent years they were named
spinons \cite{HaZi93}. The momentum of a spinon has the range
$     0 \leq k \leq \pi $
and its energy is  
\begin{equation}
 \epsilon(k) = J \pi \sin(k) \frac {\sin(\gamma)}{\gamma}
\label{spinondisp}. 
\end{equation}
The momentum and energy of an eigenstate of $H_{XXZ}$ relative to the ground
state are the sum of the momenta and energies of the individual spinons
building up the state. Two-spinon eigenstates thus yield the continuum
described by (\ref{II.8a},\ref{II.8b}). Every eigenstate of $H_{XXZ}$
belonging to the set $M_{AF}$ (see appendix)
has a fixed number of spinons. States with spin $S$ have spinon number
$N_{sp}=2S+2k, k=0,1,...$
  For finite $S$ the class C (see appendix ) of
states turns out to be identical with the set of states having the
smallest spinon number for fixed spin $S$ and maximal $S_z=S$. 
The ground state in the subspace with spin $S$ is always a class C state
with spinon number $N_{sp}=2S $.  
For $T=0$ and magnetic field $h \geq 0 $ one observes
for matrix elements of local spin operators the selection
rule, that the spinon number changes by 0 and $\pm 2$ only. This rule
is approximately valid for the Heisenberg model and exact for the
Haldane-Shastry model. 
But by far not all transition matrix elements obeying this rule are
large.  If two states differ in spinon number by 0 or 2, their energy 
difference may be much larger than $ \epsilon_2(q) $ and in that case the
matrix element is always small.

This has been observed in Ref.\onlinecite{MTBB} and confirmed by us for
larger systems. The dominant ground-state matrix elements for
magnetic field $h \geq 0$ with spin $ S $ come from two parameter continua 
of states
which are related to the ground state by the simplest hole excitations
in $\lambda$ space (the classes SWC1 and SWC2 defined in Ref.\onlinecite{MTBB}
for longitudinal correlations), whereas the dimensions of the continua
defined by $\Delta N_{sp}=0,2$ are $ 2S ,(2S+2)$ respectively. 
That means that only a small subset of all transitions with 
$\Delta N_{sp}=0,2$  contributes substantially.

\begin{figure}[h] 
\centerline{ \epsfxsize=9cm \epsfbox{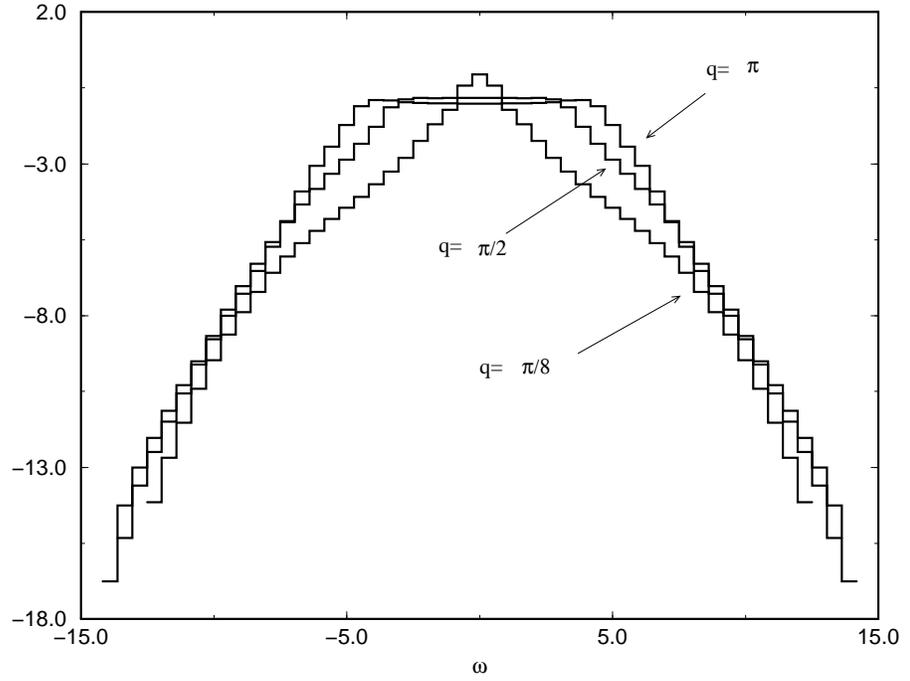} } 
\caption{ Dynamic structure factor of the $N=16$ isotropic 
Heisenberg chain: logarithm of $S^z(q,\omega)$ for $T=\infty$
and $q=\pi/8,\pi/2$ and $\pi$. }
\label{FigYY} 
\end{figure}

In the following we shall describe the nature of the excitations
relevant for $T > 0 $.  We find that $S^z(q,\omega,T)$ is for all $T$
almost
completely confined to
\begin{equation}
                      | \omega | < \varepsilon_2(q)
\label{confined}, 
\end{equation}
which is the 2-spinon boundary (\ref{II.8b}). 
The order of magnitude of suppression of the 
states lying outside of these limits can be read off 
from Fig.\ref{Fig2} for low $T$
and
from Fig.\ref{FigYY} for high $T$.

We reached this conclusion by

1. a detailed examination of a set of selected states like those from
   the $S=0$ two-spinon continuum degenerate in energy with the $S=1$
   two-spinon continuum and also higher class-C states with $S=1,2,3$
   .  In all these cases we found excitations with $\Delta N_{sp}\leq 2$ 
   and $\omega < \epsilon_2(q)$ to
   be dominant.

2. making a search in the set of all matrix elements (approximately
22.3 million ) to detect sizable contributions with exceptionally
large $\omega$ values. Not a single one was found.

The selection rules found for $T=0$ imply that only class-C states are
important : as ground states for given external magnetic field and as
excited states.  Therefore it is interesting to assess the role of
class-C states for $T>0$ too.

We computed for the $XXX$-model for $N=12$ and 16 the 
energies of all class-C states using the Bethe ansatz. To describe
the results we introduce an additional label $\alpha$ in the symbol 
for a state $ |\nu,\alpha \rangle $ where $\alpha$ indicates 
the class to which the
state belongs: a class-C state and all elements of its spin
multiplet have $\alpha=1$, a bound state (see appendix) 
has  $\alpha = 0$.
We define 
$S_{\alpha_1 \alpha_2}(q,\omega,T)$ by a formula equivalent to Eq.
 (\ref{I.5})     
with $|\nu \rangle, |\lambda \rangle$ replaced by
$|\nu,\alpha_1 \rangle, |\lambda,\alpha_2 \rangle$,
  and fixed $\alpha_1,\alpha_2$.
By integration over $\omega$
we obtain 
the four quantities $ S_{00},S_{01},S_{10},S_{11}$. The
result is shown in Table \ref{one} for $T= \infty$. It is evident
that for $T\neq 0$ and growing $N$ bound states become increasingly
important.

In Table \ref{two} we show further characteristic
quantities derived from the numerical results for
$S^z(q,\omega)$ at $T=\infty$ for $\gamma=0$ and $\gamma=0.3\pi$. 
For each $q$ we list the upper continuum boundary $\varepsilon_2(q)$
(\ref{II.8b}) and the fraction
\begin{equation}
x_{out}(q)= \frac{\int_{|\omega| > \varepsilon_2(q)} d\omega
S^z(q,\omega)}{\int d\omega S^z(q,\omega)} 
\label{xout}
\end{equation}
of the spectral weight situated outside the continuum. Even for small
$q$, where $\varepsilon_2(q)$ is small, $x_{out}(q)$ is not large, and
it becomes rapidly smaller as $q$ grows. As $q\rightarrow \pi $ the fraction
$x_{out}(q)$ becomes negligible for all practical purposes. 

The excitation continua governing the longitudinal and transversal
dynamical structure factors in the anisotropic Heisenberg model will 
be described in detail elsewhere. 

We conclude this section with results for non-Bethe-Ansatz models.
The exactly solvable Haldane-Shastry \cite{Hal88,Sha88} model
given by the Hamiltonian 
\begin{equation} 
H_{HS} =  2 J \sum_{m < n } d(m-n)^{-2} \vec S_{m} \cdot \vec S_{n}  
\label{HHS}
\end{equation}
with
\begin{equation} 
d(n) = \frac{N}{\pi} \sin \frac{\pi n }{N}
\label{chordal}
\end{equation}
describes an ideal spinon gas\cite{Hal91}.
The dynamical spin correlation functions are known exactly for $T=0$
(see Refs. \onlinecite{HaZi93,TaHa94}), but there are still 
no complete results  
for $T\neq 0$ (see Ref.\onlinecite{Tal95}).

\begin{figure}[h] 
\centerline{ \epsfxsize=9cm \epsfbox{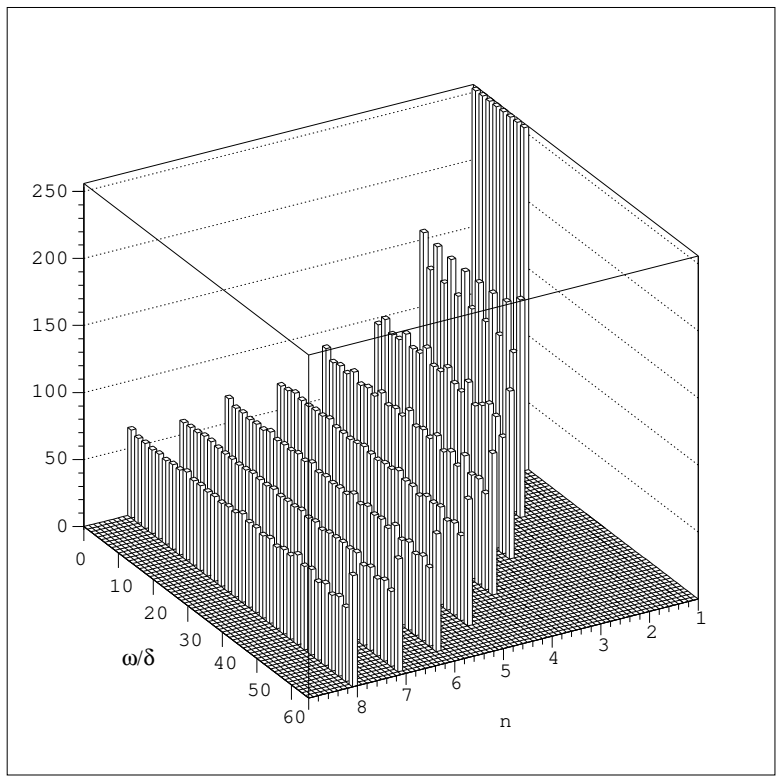} } 
\caption{The dynamic structure factor $S^z(q=\frac{n\pi}{8},\omega)$ 
for the 16-site Haldane-Shastry model, at $T=\infty$. For each of 
the equidistant excitation frequencies we plot
the sum of all squared matrix elements from (\protect{\ref{I.5}}).
Frequencies are multiples of $\delta= \pi^2/128$ .
The parabola-shaped spectral boundary as given by Haldane and Zirnbauer 
\protect\cite{HaZi93} for $T=0$ (Eq.\protect{\ref{ommax}}) is strictly 
valid for all $T$.}
\label{Fig1} 
\end{figure}

\begin{figure}[b] 
\centerline{ \epsfxsize=9cm \epsfbox{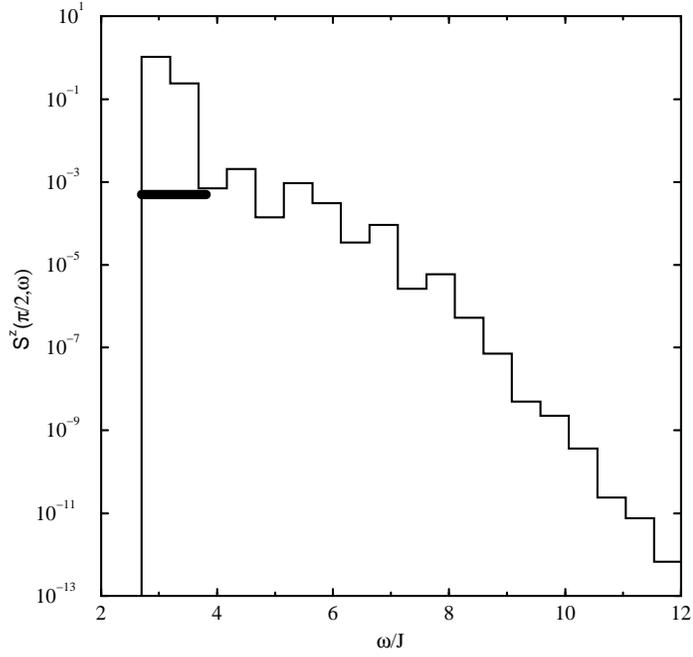} } 
\caption{
Longitudinal dynamic structure factor $S^z(\pi/2,\omega)$ at $T=0$
of the $N=16$ $XXZ$ chain with $\Delta=\cos(0.3 \pi)$.
The two-spinon continuum is marked by a horizontal bar at the
10$^{-3}$ level.
} 
\label{Fig2} 
\end{figure}

We determined for this model all transition matrix elements as given
in Eq. (\ref{I.5}) for 16 spins. Only $\omega$ which are multiples of
$\pi^2 J /{N^2}$ occur. For small chain lengths $S^z(q,\omega)$ is
far from its continuum form because the energy eigenvalues are highly
degenerate. We plot in Fig.\ref{Fig1} the sum of all
squared matrix elements for the discrete set of equally spaced
$\omega$. For $N=16$ we find that the upper limit for 
$\omega$ given by Haldane
and Zirnbauer \cite{HaZi93} for ground state excitations
\begin{equation}
                       \omega_{\rm max} = \frac{J}{2} q(2\pi-q) 
\label{ommax}
\end{equation}
is strictly valid for all matrix elements contributing to
$S^z(q,\omega,T)$. This completes observations made
by Talstra \cite{Tal95} who studied a large number of examples.

\begin{figure}[h] 
\centerline{ \epsfxsize=9cm \epsfbox{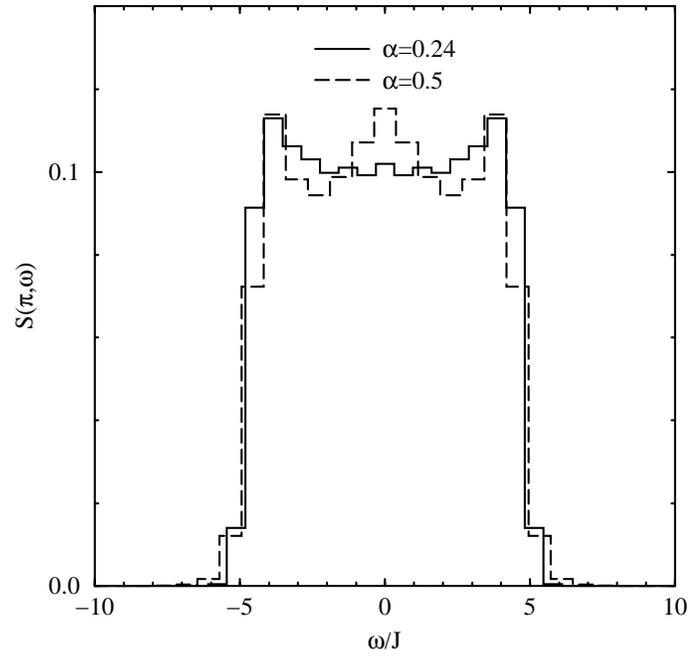} } 
\caption{The dynamic structure factor $S^z(\pi,\omega)$ 
for the 16-site frustrated Heisenberg
model with $\alpha = 0.24$ (solid line) and $\alpha = 0.5$ (dashed line), at
$T=\infty$.}
\label{Fig3} 
\end{figure}

Finally we discuss results for another non-Bethe-Ansatz model.
To explain the properties of quasi-one-dimensional compounds
the description in terms of a spin model including interactions
between both nearest and next-nearest-neighbors may be
necessary. The frustrated model

\begin{equation}
H_{\rm frus} = 2 J \sum_{i = 1}^{N} \left\{ \vec S_{i} \cdot \vec S_{i + 1}  
+ \alpha \vec S_{i} \cdot \vec S_{i + 2}  
\right\}  
$$
\label{Hfrus},
\end{equation}
has been used in attempts to describe experimental data for
CuGeO$_3$ with $\alpha=0.24$ \cite{CEC95} and Sr$_{14}$Cu$_{24}$O$_{41}$
with $\alpha=0.5$ (see Ref.\onlinecite{MK96}).
It is therefore interesting to find out whether the spin dynamics
is significantly changed when additional interactions are introduced.
In Fig.\ref{Fig3} we compare $S^z(q=\pi,\omega,T=\infty)$ for 
two $\alpha$ values.
We see that the spectral weight vanishes outside the two-spinon
boundary of the isotropic Heisenberg model for both models. 

\subsection{Results for $S^{\mu}(q,\omega)$}
\label{VB}

In this section we describe the results for the dynamic structure
factors $S^{\mu}(q,\omega)$ for $H_{XXZ}$ (\ref{I.1}).  For any finite
$N$, $S^{\mu}(q,\omega)$ (as given in Eq.(\ref{I.5})) 
consists of a finite number of
delta peaks.  Of course the number of peaks grows exponentially with
$N$, but only a rather small fraction of them
contributes significantly. The number of these relevant peaks varies
strongly with $N$, as do their precise locations and strengths.  (The
regular structure visible in Fig.\ref{Fig1} for the Haldane-Shastry
model is a striking exception to this rule.)  In order to avoid
confusion by these finite-size effects we present our results for
$S^{\mu}(q,\omega)$ as histograms, displaying for each frequency bin
of constant size $\Delta \omega$ the total spectral weight of the
$S^{\mu}(q,\omega)$ peaks in the bin, divided by $\Delta \omega$.  The
size of the frequency bins chosen represents a compromise: too narrow
bins overemphasize finite-size effects, too wide bins erase all
structure.

\subsubsection{The $XX$ model}
\label{VB0}

\begin{figure}[b] 
\centerline{ \epsfxsize=9cm \epsfbox{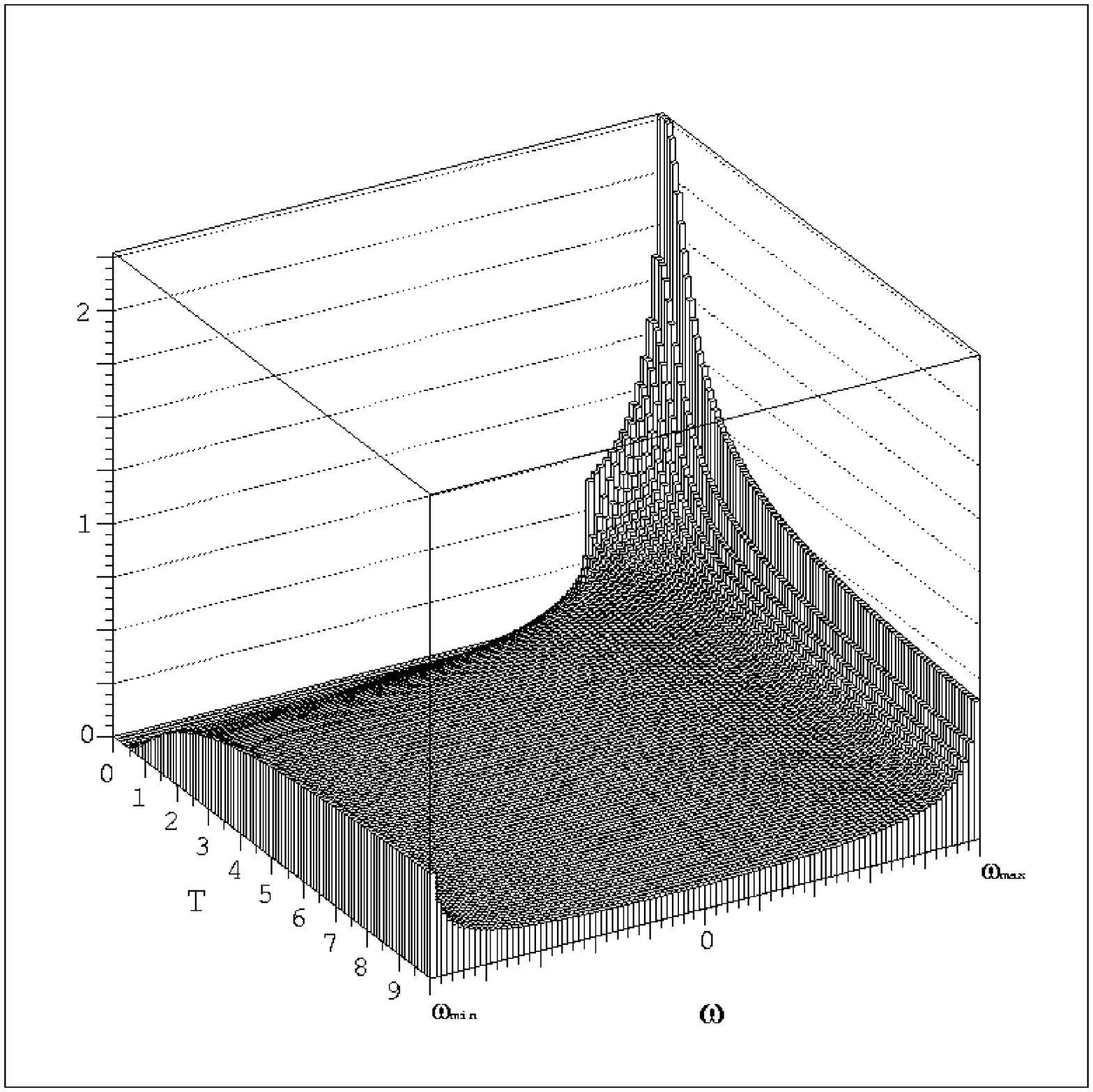} } 
\caption{Longitudinal dynamic structure factor $S^z(\pi/2,\omega,T)$
of the $XX$ chain. (See Eq. (\ref{II.6}).) Note the sharp cutoff
at the lower spectral boundary near $T=0$. } 
\label{Fig4} 
\end{figure}

For the $XX$ model, $S^{z}(q,\omega)$ is known in closed form (see
\ref{II.6}), but $S^{x}(q,\omega)$ is only known numerically, except
at infinite $T$ (see \ref{II.12}).  The shape of $S^{z}(q,\omega)$ 
(see Fig.\ref{Fig4})  is
dominated for low $T$ by the spectral boundaries
at $\varepsilon_1$ and $\varepsilon_2$ and in general 
by the inverse-square-root singularities 
$\omega= \pm \varepsilon_2(q)$ (\ref{II.8b}). For finite
$T$ the $\omega$-dependence of $S^{z}(q,\omega)$ thus shows a typical
``trough'' shape (see Fig.\ref{Fig4}). This trough shape is a 
feature which is present also in the $XXZ$ model
results to be discussed below.

\begin{figure}[h] 
\centerline{ \epsfxsize=9cm \epsfbox{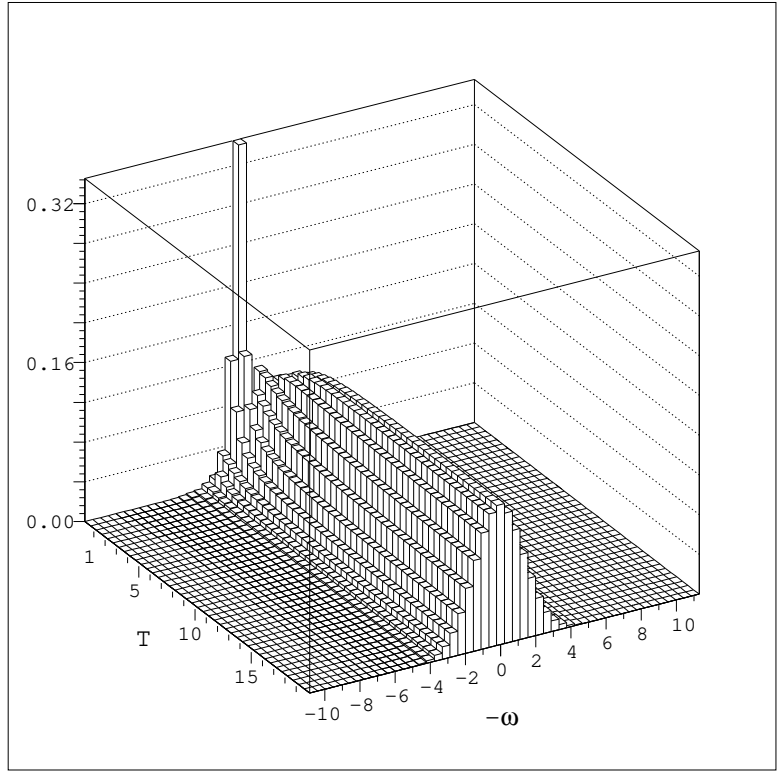} } 
\caption{ Transversal dynamic structure factor 
$S^x(q=\pi/2,\omega,T)$. The $T=const$ cross section  
becomes an exact Gaussian for $T=\infty$. } 
\label{Fig5} 
\end{figure}

The function $S^{x}(q,\omega)$ as shown in Fig.\ref{Fig5} 
involves in the Jordan-Wigner fermion picture the excitation 
of arbitrarily many particle-hole pairs and
thus its frequency range is not restricted. This is reflected
in the $q$-independent Gaussian shape of $S^{x}(q,\omega)$ at
$T=\infty$: spectral weight is present at all frequencies, but it is
very small for high frequencies. The ``hill'' shape exemplified by the
Gaussian again is a generic feature which persists also for $\gamma
\neq 0.5 \pi$.

At $\gamma = 0.5 \pi$ we find the typical hill structure in
$S^{x}(q,\omega)$ for all $q$ and for all $T \gtrsim 2 J$. At low $T$
 the bulk of the spectral weight is contained in the region
confined by the continuum boundaries $\varepsilon_1(q),
\varepsilon_2(q)$, and $\tilde{\varepsilon}_2(q)$ conjectured
\cite{MTPB81} for $T=0$. 

\subsubsection{The $XXZ$ model, low T}
\label{VCa}

Here we discuss our low-temperature results, starting
with the longitudinal structure factor. 
As discussed in Sec. \ref{VA},
the low-temperature $N=16$
data for $\gamma=0$ show that the spectral weight in $S^z(q,\omega)$ 
is almost completely confined to the continuum
between $\varepsilon_1(q)$ and $\varepsilon_2(q)$.

Recently it was shown \cite{BCK96,KMB96} that the two-spinon
contribution to $S^{x,z}(q,\omega)$ vanishes in a square-root cusp at
the upper boundary.

The contributions to $S^{\mu}(q,\omega)$ from outside the two-
spinon continuum become visible on a logarithmic scale. 
In Fig.\ref{Fig2} we have plotted $S^z(\pi/2,\omega)$ for 
$\Delta=\cos(0.3 \pi)$ and $N=16$. The two frequency bins 
containing almost all of the spectral weight lie 
within the boundaries of the two-spinon continuum; beyond 
$\varepsilon_2(q)$, 
$S^z(\pi/2,\omega)$ decays rapidly.

 At low $T$ the continuum boundaries are generally quite well visible
in the numerical results for both $\gamma=0$ and 0.3$\pi$. (Compare
Fig.\ref{Fig6} for $\gamma=0.3 \pi$ and Fig.\ref{Fig8} for
$\gamma=0 ; q=3 \pi/4$ in both cases.)  At $\gamma=0$ and small $q$,
there is some spectral weight rather far above the upper continuum
boundary $\varepsilon_2(q)$. The maximum of $S^z(q,\omega)$ is close
to the lower continuum boundary for those $q$ values where the
continuum boundaries are well separated on the scale given by the
frequency bin width. For $\gamma=0$ that is consistent with the
conjecture of Ref.\onlinecite{MTPB81}.

The low-temperature results for the transverse structure factor
 $S^x(q,\omega)$  at $\gamma=0.3 
\pi$ (compare Fig.\ref{Fig7} for $q=3 \pi/4$)
are similar to those at $\gamma=0.5 \pi$. The continuum 
boundaries are again quite well recognizable, with only very 
little spectral weight above the upper continuum boundary. The 
maximum of  $S^x(q,\omega)$ is always situated close to the lower 
spectral boundary and its height grows with $q$. 

We have carried out a detailed comparison between Schulz'
field-theoretical result (\ref{S1}) and our numerical $N=16$ results
for $S^{\mu}(q,\omega)$ 
at $q=7 \pi /8$ and $\pi$. As expected, (\ref{S1})
works best in the low-energy region at low temperatures. For $T
\rightarrow 0$ the singularities show up at the lower boundary 
$\varepsilon_1(q)$ of the continuum. There are no sharp high-frequency 
cutoffs nor singularities, but apart from that, the shapes 
of the two curves are similar. As an example we show in Fig.\ref{Fig9}
$S^z(7\pi/8,\omega) $ for the $XXX$ model at
$T=1.2J$ (see Ref.\onlinecite{constfix}). There the low-frequency data 
are described well by (\ref{S1}), and the high-frequency cutoff is replaced by
a gradual but quite rapid decay.

As the temperature is raised
to $T = 2.7 J$ (as seen in Fig.\ref{Fig10}) Eq. (\ref{S1})
extends further beyond the range of the numerical data for both positive
and negative frequencies. The maximum of  (\ref{S1}) has moved towards
higher frequencies as compared to Fig.\ref{Fig9} and as $T$ is
raised further the maximum moves to frequencies where the
numerical data essentially vanish.

The result (\ref{S1}) of Schulz was successfully \cite{TCNT95} 
used to fit the $T$-dependence of inelastic neutron scattering
results on the Heisenberg antiferromagnetic chain compound KCuF$_3$ in
the vicinity of $q=\pi$. We wish to point out that the agreement with
the experimental data found in that study does not contradict the
differences to our numerical results described above. Firstly, the
experiments were all carried out in the range $ T \lesssim J$ where
(\ref{S1}) is very good at low frequencies and not unreasonable at
high frequencies.  Secondly, (\ref{S1}) was only used to describe the
experimental data in the region $q \approx \pi$ at frequencies far
below the upper continuum boundary $\varepsilon_2(q)$. High-frequency
data (see, for example, Fig.17 of Ref.\onlinecite{TCNT95}) were
interpreted in terms of the $T=0$ conjecture of
Ref.\onlinecite{MTPB81}, and they did show a clear spectral cutoff
consistent with that conjecture (and thus also with our numerical
data).

\subsubsection{The $XXZ$ model, high T}
\label{VCb}
The temperature dependence of the dynamic structure factor displays
clear general trends: For small $q$, $S^{\mu}(q,\omega)$ is largest at
high $T$, whereas at larger $q$, $S^{\mu}(q,\omega)$ decreases with
$T$.  For $\Delta \neq 0$, the high-temperature dynamic structure
factor $S^{\mu}(q,\omega)$ always has its maximum at small $q$. 
Given that $S^z(q,\omega,T)$ is almost completely confined to
$|\omega| < \varepsilon_2(q)$ this is a consequence of the
sum rule
\begin{equation}
\int d\omega S^z(q,\omega,T=\infty)=\pi/2
\label{sumrule}.
\end{equation}

The general shape of the high-temperature structure factor 
$S^{\mu}(q,\omega)$  (for $N=16$) changes smoothly as one 
goes from $\gamma=0.5\pi$ to
$\gamma=0$: at $\gamma=0.3\pi$ $S^{x}(q,\omega)$ shows the ``hill''
shape familiar from $\gamma=0.5\pi$, and  $S^{z}(q,\omega)$ shows the
``trough'' shape. At $\gamma=0$, $S^{z}(q,\omega)$
 is of ``hill'' type for $q < \pi/2$ and of
``trough'' type for $q > \pi/2$. 

\begin{figure}[h] 
\centerline{ \epsfxsize=7cm \epsfbox{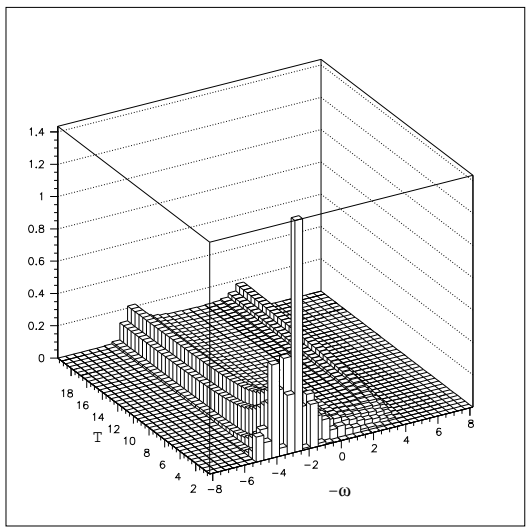} } 
\caption{ Temperature dependence of the
longitudinal dynamic structure factor $S^z(3\pi/4,\omega)$ 
of the $N=16$ $XXZ$ chain with $\Delta=\cos(0.3 \pi)$.
$T$ varies from $0.2 J$ (front) to $19.7 J$ (back). The
two-spinon continuum ranges from $\omega=1.907 J$ to
$\omega=4.983 J$. Note that the $x$ axis represents $-\omega / J$ for
reasons of visibility.} 
\label{Fig6} 
\end{figure}

\begin{figure}[h] 
\centerline{ \epsfxsize=7cm \epsfbox{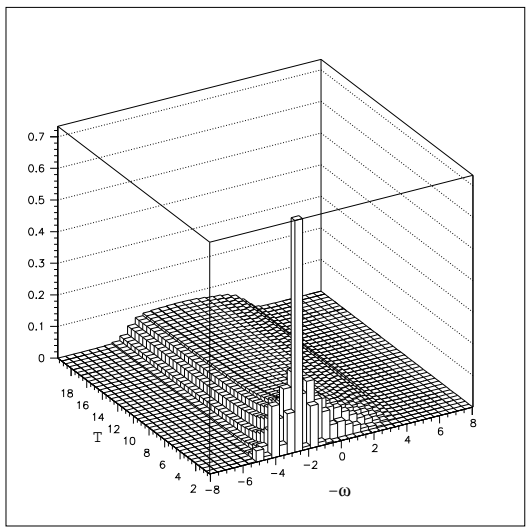} } 
\caption{ Same as Fig.\protect{\ref{Fig6}}, for the transverse
dynamic structure factor $S^x(3\pi/4,\omega)$. The boundaries of the
excitation continua are $\varepsilon_1=1.907 J, \varepsilon_2=4.983 J,$
and $\tilde{\varepsilon}_2=2.064 J$. The $x$ axis
 represents $-\omega / J$ for reasons of visibility.
} 
\label{Fig7} 
\end{figure}

\begin{figure}[h] 
\centerline{ \epsfxsize=7cm \epsfbox{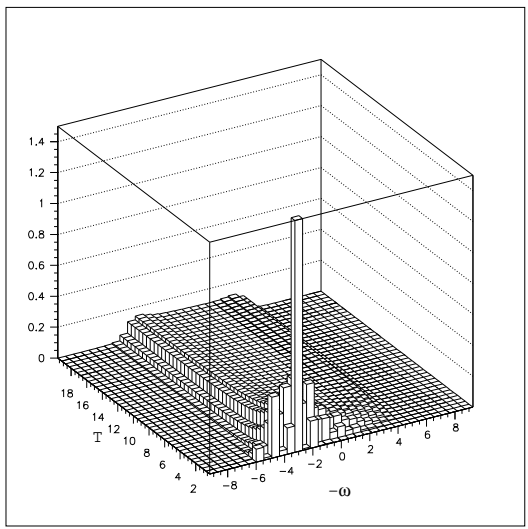} } 
\caption{ Temperature dependence of the
dynamic structure factor $S^z(3\pi/4,\omega)$ 
of the $N=16$ isotropic Heisenberg chain.
$T$ varies from $0.2 J$ (front) to $19.7 J$ (back). 
The two-spinon continuum ranges from $2.22 J$  to $5.803 J$. The $x$ axis
represents $-\omega / J$ for reasons of visibility.
} 
\label{Fig8} 
\end{figure}

\begin{figure}[h] 
\centerline{ \epsfxsize=9cm \epsfbox{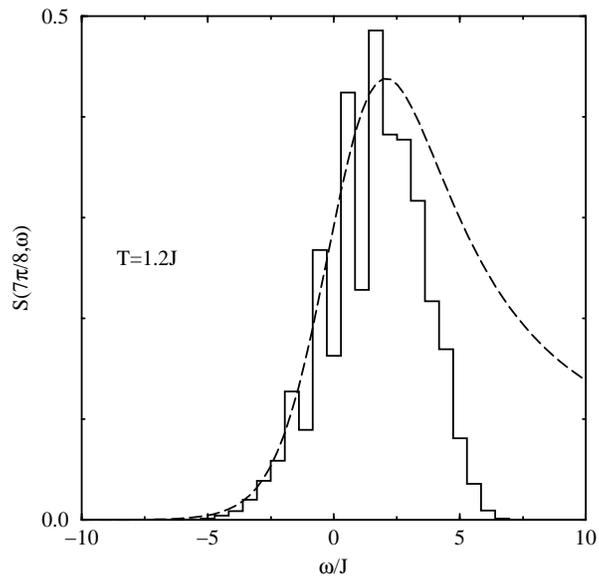} } 
\caption{Dynamic structure factor 
$S^z(7\pi/8,\omega)$ of
the $XXX$-chain at $T=1.2J$.
Shown are the numerical result from a $N=16$ chain (solid line) and the
field-theoretical result of Schulz, Eq. (\protect\ref{S1}) (dashed line).
} 
\label{Fig9} 
\end{figure}

\begin{figure}[h] 
\centerline{ \epsfxsize=9cm \epsfbox{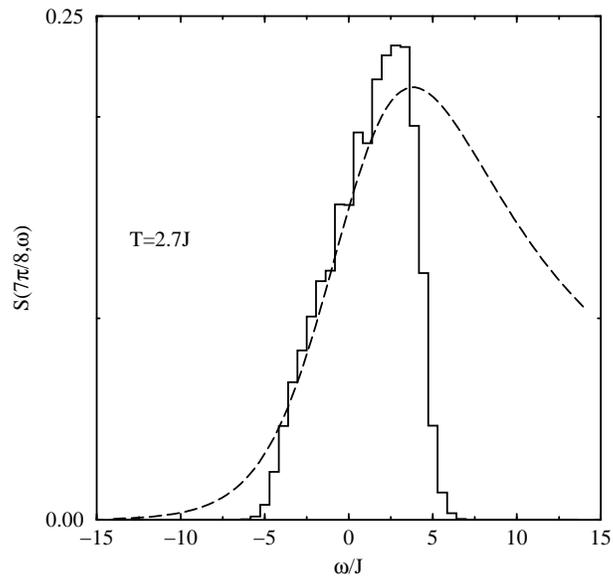} } 
\caption{ 
Same as Fig.\protect\ref{Fig9}, for $T=2.7J$.
} 
\label{Fig10} 
\end{figure}

Clearly, the number of terms 
contributing to $S^{\mu}(q,\omega)$ (\ref{I.5}) at $T\neq0$ is much
larger than at $T=0$,
and as $T \rightarrow \infty$, the contributions 
of different states $|\nu \rangle$ in (\ref{I.5}) are not 
discriminated by the thermal weight factor $\exp(-\beta E_{\nu})$ 
any more. 
Nevertheless Figs.\ref{Fig5}-\ref{Fig8} show that $S^{\mu}(q,\omega,T)$ is
confined to a narrow interval in $\omega$, as discussed in Sec. \ref{VA}. 
In Fig.\ref{Fig6} we show $S^z(\frac{3 \pi}{4}, \omega)$ 
at $\gamma=0.3 \pi$ for a range of temperatures. The continuum boundaries 
(\ref{II.8a}) and (\ref{II.8b}) are $\varepsilon_1=1.906J$ and 
$\varepsilon_2=4.982J$. For low $T$,  $S^z(\frac{3 \pi}{4}, 
\omega)$ is basically confined to the interval between  
$\varepsilon_1$ and  $\varepsilon_2$ and shows sharp structures 
(probably due to finite-size effects) and a maximum close to  
$\varepsilon_1$. For higher $T$ these structures are washed out 
and the probability of negative-frequency contributions to  
$S^z(\frac{3 \pi}{4}, \omega)$ grows in accordance with detailed
balance $S^{\mu}(q,-\omega,T)= S^{\mu}(q,\omega,T) \exp(-\beta \omega )$ . No 
significant spectral weight appears for $|\omega| > \varepsilon_2$ 
as $T$ grows. 

\begin{figure}[h] 
\centerline{ \epsfxsize=9cm \epsfbox{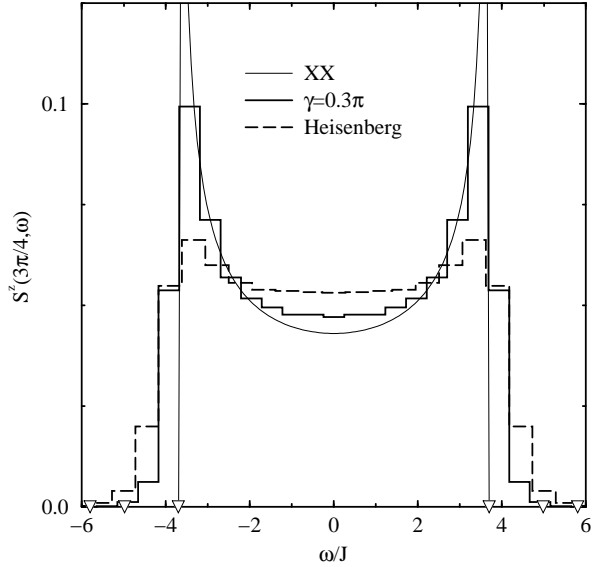} } 
\caption{Longitudinal dynamic structure factor $S^z(3\pi/4,\omega)$ 
of the $N=16$ $XXZ$ chain at $T=\infty$, for $\Delta=1$ (dashed line)
and $\Delta=\cos(0.3 \pi)$ (solid line).
For comparison we also show the exact $N=\infty$ result for 
the $XX$ chain, $\Delta=0$ (thin line). The three pairs of
triangles on the frequency axis mark the upper boundary $\pm
\varepsilon_2(\frac{3 \pi}{4})$ (\protect\ref{II.8b}) of the two-spinon
continuum for $\Delta=0, \cos(0.3 \pi)$, and 1 (in ascending order of
$|\omega|$). 
} 
\label{Fig11} 
\end{figure}

Fig.\ref{Fig11} shows $S^z(\frac{3 \pi}{4}, 
\omega)$  at $T=\infty$, for $\gamma=0, 0.3\pi$ (for $N=16$) and 
$\gamma=0.5 \pi$ for $N=\infty$. The 
maxima of $S^z(\frac{3 \pi}{4}, \omega)$  for $\gamma=0$ and $0.3 
\pi$ coincide with the singularities for the case, $\gamma=0.5 
\pi$. For frequencies 
beyond the  maxima (where $S^z(\frac{3 \pi}{4}, \omega)$ vanishes
strictly for 
$\gamma=0.5 \pi$) we see a sharp decline for $\gamma=0$ and $0.3 \pi$ 
with small tails extending outward, roughly to   $\varepsilon_2 
(\frac{3\pi}{4},\gamma)$, where  $S^z(\frac{3 \pi}{4},\omega)$ 
becomes negligible on a linear scale. (On a logarithmic scale, 
rapidly decaying contributions for large $|\omega|$ are visible, 
similar to those in Fig.\ref{Fig2}; see also Fig.\ref{Fig13}.) 
The observations from Fig.\ref{Fig11} illustrate the data
presented in Table \ref{two}. 

\begin{figure}[h] 
\centerline{ \epsfxsize=9cm \epsfbox{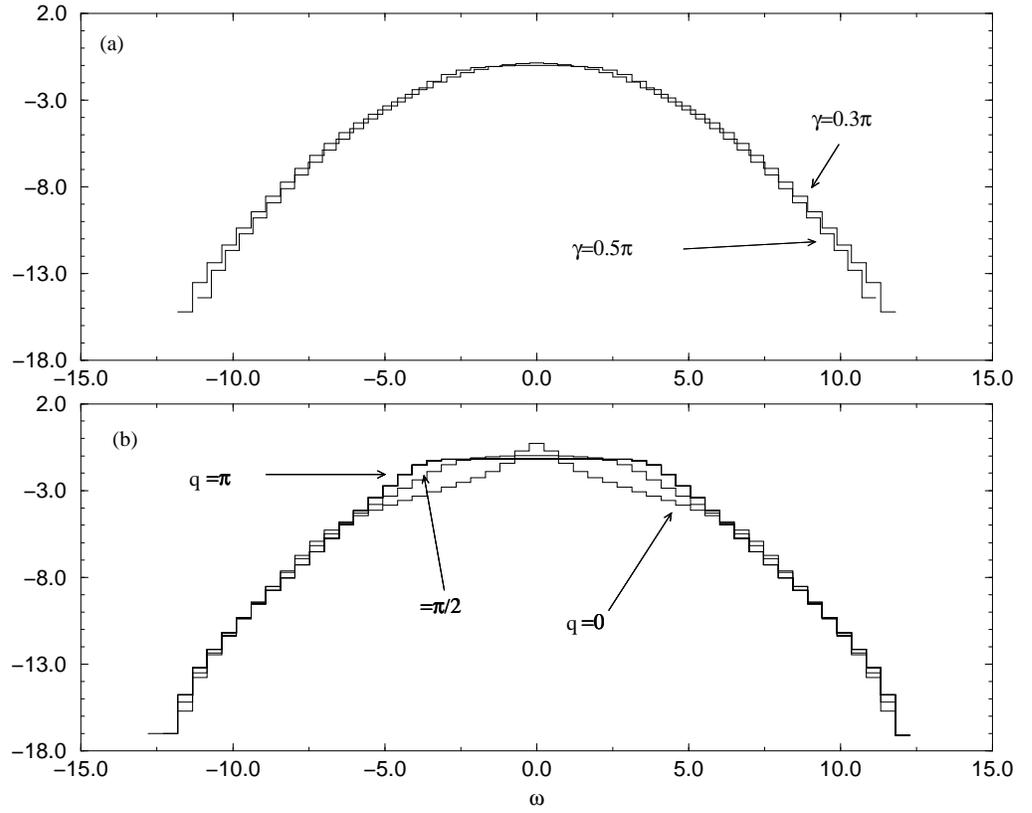} } 
\caption{Transversal dynamic structure factor: logarithm 
of $S^x(q,\omega)$ of the $N=16$ $XXZ$-chain 
with $\gamma=0.3\pi$ and $\gamma=0.5\pi$ at $T=\infty$.
a: $q=\frac{\pi}{2}$ for $\gamma=0.3 \pi$ and the q-independent
Gaussian for $\gamma=0.5\pi$.
b: $\gamma=0.3\pi$ and $q=0,\pi/2$ and $\pi$. }
\label{FigXX} 
\end{figure}

The transversal dynamical structure factor 
$S^x(q,\omega,T=\infty)$ evolves 
for varying anisotropy (from $\gamma=0$ to $\pi/2$) 
from a strongly $q$-dependent 
function with  spectral boundaries $\pm\epsilon_2(q)$ 
(see (\ref{II.8b}))
to the $q$-independent Gaussian Eq. (\ref{II.12}) (see 
Figs. \ref{FigYY},\ref{FigXX} for details). There is no physical
explanation for this change of shape yet.

\subsection{Some remarks on finite-size effects}
\label{VD}

In contrast to 
the case of static structure factors, where systematic 
finite-size analyses can be performed at both zero and finite $T$, 
the presence of an additional variable makes the situation more 
complicated for dynamic 
structure factors. As stated at the beginning of this section, 
the number, the weights and the positions of the $\delta$ 
functions contributing to $S^{\mu}(q,\omega)$ (\ref{I.5}) vary
with the system size $N$ in a complicated manner. In the 
histogram representation which we use this means that as $N$ 
varies, spectral weight is exchanged between neighboring bins in 
a seemingly random manner. 

\begin{figure}[h] 
\centerline{ \epsfxsize=9cm \epsfbox{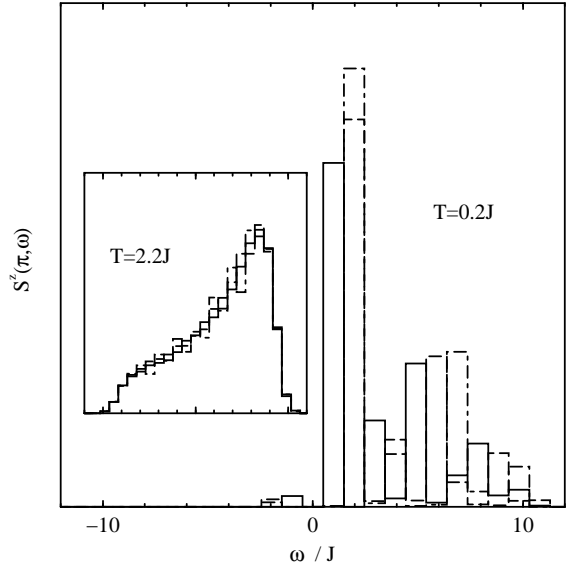} } 
\caption{ Finite-size effects in $S^z(\pi,\omega)$ for 
$\gamma=0.3 \pi$  at $T=0.2 J $ (main plot) and $T=2.2 J$ 
(inset). Data for $N=12, 14,$ and 16 are shown as dot-dashed, 
dashed, and solid lines, respectively.
} 
\label{Fig12} 
\end{figure}

\begin{figure}[h] 
\centerline{ \epsfxsize=9cm \epsfbox{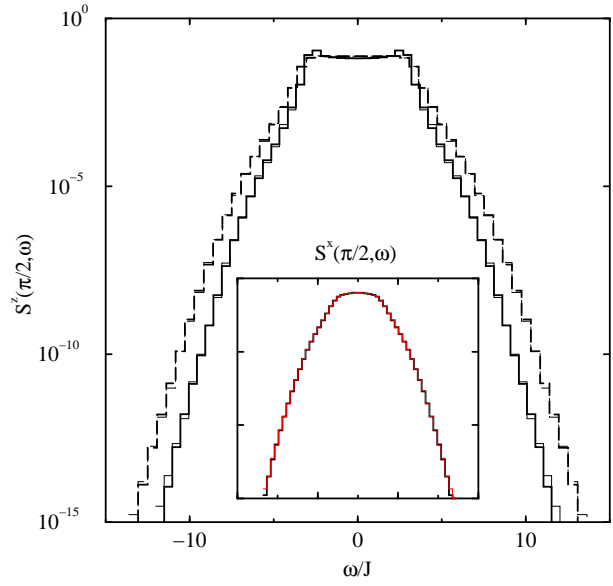} } 
\caption{Absence of finite-size effects at $T=\infty$ in the
low-weight spectral tails. Main plot: $S^z(\pi/2,\omega)$ at
$\gamma=0.3 \pi$ (solid line) and $\gamma=0$ (dashed line). Inset:
$S^x(\pi/2,\omega)$ at $\gamma=0.3 \pi$(same range of variables as main
plot). Data for $N=12$ are shown as thin lines, those for $N=16$ as
heavy lines.  }
\label{Fig13} 
\end{figure}

For the rather wide bins used here and for high $T$, 
where many states $|\nu \rangle$ in (\ref{I.5}) contribute
the weight exchanged between two bins is very small compared 
to the total weight in each bin and thus finite-size 
effects are small.
At low $T$, where a small number of $\delta$ peaks dominates  
$S^{\mu}(q,\omega)$ we do indeed observe larger finite-size effects. 
This is seen in Fig.\ref{Fig12} (main plot), where we show
$S^z(\pi,\omega)$ at $T=0.2J, \gamma=0.3 \pi$ for $N=12, 14,$ and 
16. The same quantity, at $T=2.2J$ is shown in the inset of 
Fig.\ref{Fig12} demonstrating that
with the general smoothing effect of nonzero $T$, the finite-size 
effects in  $S^{\mu}(q,\omega)$ also become smaller rapidly. 
The rapidly decaying spectral
tails in $S^{\mu}(q,\omega)$ which extend well beyond the two-spinon
continua show only very little size-dependence at $T=\infty$, as 
shown in Fig.\ref{Fig13} for $q=\pi/2$ at $\gamma=0$ and $0.3 
\pi$.

\section{Time-dependent correlation functions}
\label{III}

\begin{figure}[h] 
\centerline{\epsfxsize=9cm \epsfbox{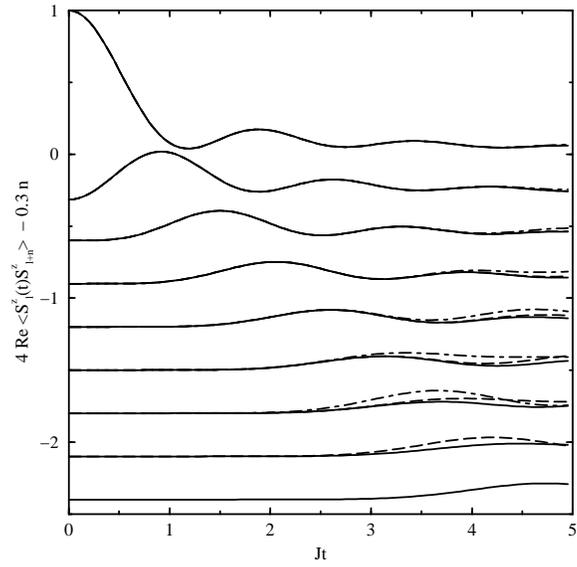} } 
\caption{Real part of the $z$ spin
pair correlation function $\langle S_l^{z}(t) S_{l+n}^{z} \rangle$ for
$n=0$ (top) through 8 (bottom), at $T/J=20$, for anisotropy
$\Delta=\cos(0.3 \pi)$ and different system sizes, $N \geq 2n$, namely
$N=16$ (solid lines), $N=14$ (dashed), and $N=12$ (dot-dashed). (Curves
are displaced in the vertical direction for graphical purposes.)}   
\label{Fig14}
\end{figure}

In Fig.\ref{Fig14} we show the real part of $\langle S_l^{z}(t) S_{l+n}^{z}
\rangle$ for $n=0,1,\ldots,8$ at $T/J=20$, for $N=12, 14,$ and 16, in
the $XXZ$ chain with $\Delta= \cos(0.3 \pi)$. As in all plots of this
type 
the results for
different $N$ are identical at short times and differ from
each other at longer times in a systematic way, with the data for the
smallest $N$ deviating first.
(In the $XX$ case
this
behavior can be observed over a wide range of system sizes.) Thus
the results are certainly valid for $N=\infty$ over the
time range where $N=12, 14$, and 16 yield identical correlations. The
onset of finite-size effects occurs earlier for larger $n$. This is
easily interpreted by observing that the finite-size effects are due
to correlations connecting the sites $l$ and $l+n$ ``the wrong way
round'', i.e. over a distance $N-n$ instead of n. The (roughly) linear
growth with $n$ of the time range affected by finite-size effects is
consistent with ballistic propagation of the correlations. In an
analogous way one may understand the linear growth with $n$ of the
short-time region of practically vanishing correlations: the
information from site $l$ has not yet arrived at site $l+n$.

\begin{figure}[b] 
\centerline{ \epsfxsize=9cm \epsfbox{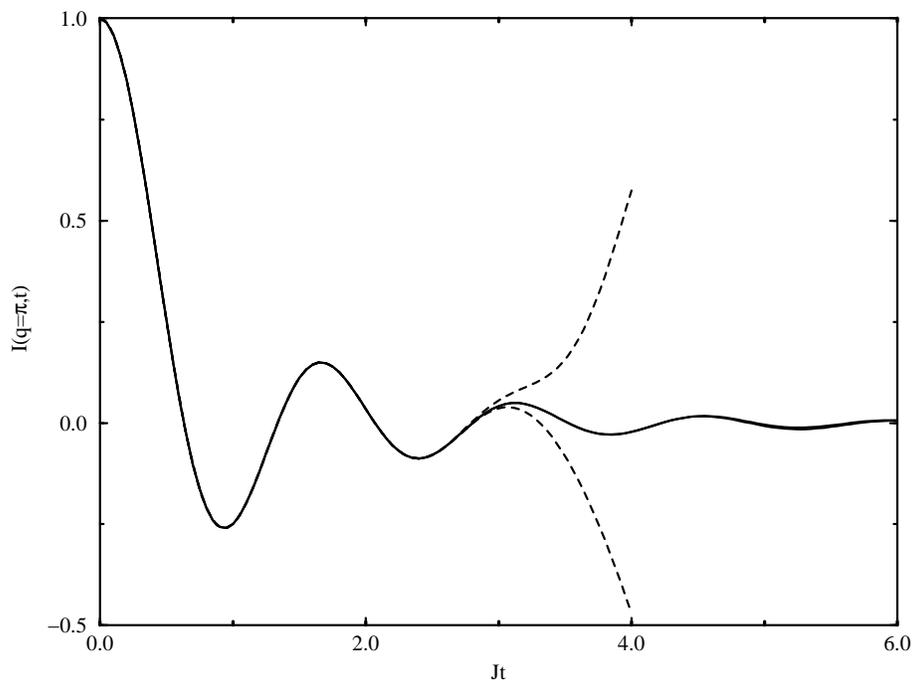} } 
\caption{The intermediate structure factor 
$I^z(q=\pi,t,T=\infty)$ for the $XXX$-chains of 14 and 16 spins and the upper
and lower boundaries (dashed lines) from Ref.\protect\onlinecite{BL92}.
The curves for $N=14$ and $N=16$ can not be distinguished on this
scale.}
\label{Fig15} 
\end{figure}

At $T=\infty$, we compared the finite-$N$ data to exact 
infinite-$N$ results \cite{BL92} obtained by moment-expansion 
methods. These methods employ the power-series expansion 
coefficients of 
$\langle S_{l+n}^{\mu}(t) S_l^{\mu} \rangle$
(which is an even real function of $t$ at $T=\infty$) up to a 
certain  order to obtain exact upper and lower bounds to 
$\langle S_{l+n}^{\mu}(t) S_l^{\mu} \rangle$
(or to $I^{\mu}(q,t)$) for the infinite system. Over a certain 
time range the upper and lower bounds coincide (for all practical 
purposes) and then they start to deviate from each other rapidly. 
B\"ohm and Leschke  \cite{BL92} obtained the expansion 
coefficients to order $t^{30}$ for the $XXX$-chain 
and the resulting bounds coincide with each other 
(and thus with the exact $ N=\infty $ result) for $Jt \lesssim 2.8$. 
In Fig.\ref{Fig15} we show these bounds together with our results for
$N=14$ and $N=16$.

\begin{figure}[h] 
\centerline{ \epsfxsize=9cm \epsfbox{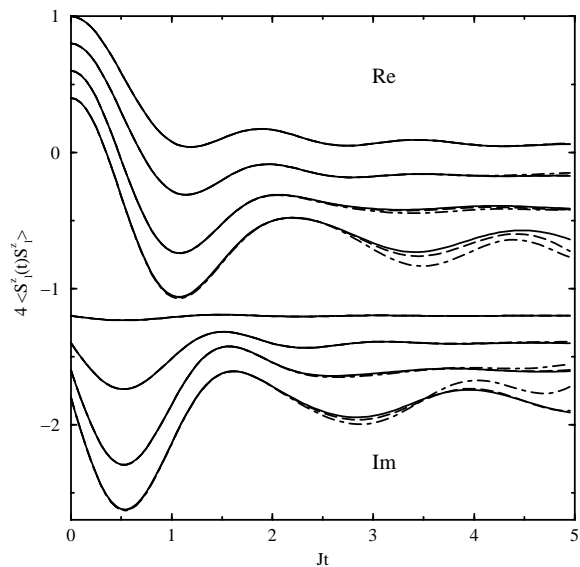} } 
\caption{Real part (upper set of curves) and
imaginary part (lower set of curves) of the $z$ spin autocorrelation
function $\langle S_l^{z}(t) S_{l}^{z} \rangle$ for temperatures
$T/J=20, 2, 0.7,$ and $0.2$ (top to bottom in each set of curves),
for anisotropy $\Delta=\cos(0.3 \pi)$ and different system sizes,
$N=16$ (solid lines), $N=14$ (dashed), and $N=12$ (dot-dashed). (Curves
are displaced in the vertical direction for graphical purposes; the
$y$-axis scale applies to the uppermost curve; Re$\langle S_l^{z}(0)
S_{l}^{z} \rangle = 1$, Im$\langle S_l^{z}(0) S_{l}^{z} \rangle = 0$
for arbitrary $T$.)}
\label{Fig16} 
\end{figure}

In Fig.\ref{Fig16} we display the $z$ spin
autocorrelation function $\langle S_l^{z}(t) S_{l}^{z} \rangle$ for
$\Delta= \cos(0.3 \pi)$ and four representative $T$ values. Two
general tendencies can be observed in these data as $T$ is lowered:
firstly, the oscillations in the correlation function become more
pronounced, and secondly, finite-size effects start to show up earlier
at lower $T$, due to the small (size-dependent) 
density of states at low frequencies. 
The dominant time scale of the oscillations, however,
does not change with $T$.

\begin{figure}[h] 
\centerline{ \epsfxsize=9cm \epsfbox{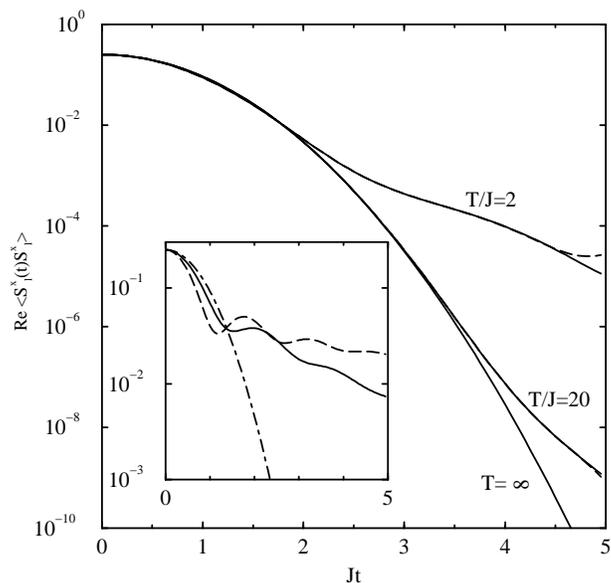} } 
\caption{Real part of the $x$ spin autocorrelation 
function $\langle S_l^{x}(t) S_{l}^{x} \rangle$ of the $XX$ chain 
$(\Delta=0)$ for temperatures $T/J=2$, $20$, and $\infty$, and system sizes 
$N=16$ (solid lines) and $N=14$ (dashed lines; no $N=14$ data shown
for $T=\infty$). The $N=16$ result for  $T=\infty$ coincides with
the exact result for  
$N=\infty$ following by Fourier transformation from 
(\protect\ref{II.12}). The inset shows the $N=16$ results
for $T=\infty$, $\Delta=0$ (dot-dashed),  $\Delta=\cos(0.3\pi)$ (solid
line),  and 
$\Delta=1$ (dashed line).}
\label{Fig17} 
\end{figure}

In Fig.\ref{Fig17} (main plot) we show the real part of the transverse
autocorrelation function $\langle S_l^{x}(t) S_{l}^{x} \rangle$ of the
$XX$ chain for three (high) temperature values, in
comparison to the exact Gaussian infinite-$T$ (and infinite-$N$) result.
At short times, all results coincide. The $T/J=2$
results branch off first, the $T/J=20$ results follow later. The
$T=\infty$ results for $N=16$ coincide \cite{note1} with the Gaussian
over the time range shown here. In the
non-Gaussian regime (for finite $T$), 
the data for N=16 and 14 stay together for a while,
 before finite-size effects become important. The short 
time interval
between the Gaussian and finite-size dominated regimes displays the
exponential decay predicted \cite{IIKS93} and verified \cite{SNM95}
recently, with the expected value of the decay rate.  
The inset of Fig.\ref{Fig17} demonstrates the changes that occur
in the infinite-$T$ $x$ autocorrelation when the interaction $\Delta$ is
switched on. A finite $\Delta$ both accelerates the initial decay of
$\langle S_l^{x}(t) S_{l}^{x} \rangle$ (an effect well known from
short-time expansion studies at $T=\infty$ \cite{RMP86,BL92}) and
decelerates the long-time decay, which becomes much slower than the
exponential decay present at $\Delta=0$. 
This observation is evidence for the exceptional character of the
high-temperature  $x$ autocorrelation function at $\Delta=0$. The
limited time range covered by our data prevents a 
quantitative characterization of the changes induced in the long-time 
behavior by a nonzero $\Delta$ \cite{note2}.

\begin{figure}[h] 
\centerline{ \epsfxsize=9cm \epsfbox{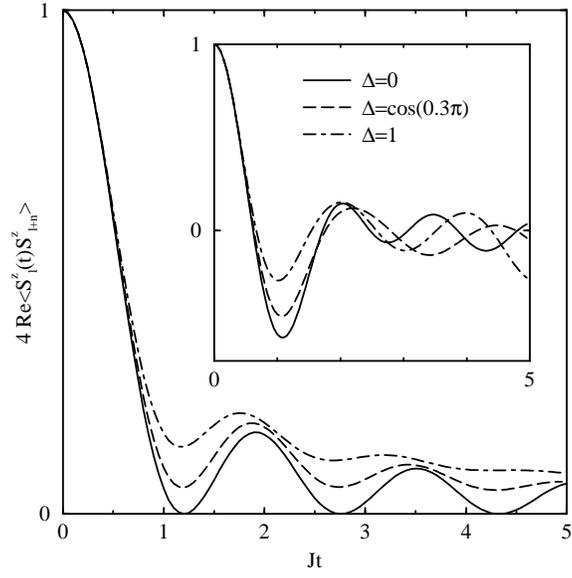} } 
\caption{Real part of the $z$ spin autocorrelation
function $\langle S_l^{z}(t) S_{l}^{z} \rangle$ of the $XXZ$ chain
for $N=16$ and temperatures $T/J=\infty$ (main plot) and $0.2$ 
(inset).}
\label{Fig18} 
\end{figure}

In contrast to the rather pronounced slowing-down of the long-time
 decay of $\langle S_l^x(t) S_l^x \rangle$ with growing $\Delta$ at
 high $T$ (see Fig.\ref{Fig17} (inset)), the long-time
 decay of $\langle S_l^{z}(t) S_{l}^{z} \rangle$ varies only weakly
 with $\Delta$.  This is demonstrated in Fig.\ref{Fig18} (main
 plot) for $T=\infty$: the zeros of $\langle S_l^z(t) S_l^z \rangle$
 at $\Delta=0$ (compare (\ref{II.1})) gradually
 ``fill up'' as $\Delta$ grows, but the overall decay stays roughly
 the same. 
The curves for Re$\langle S_l^z(t) S_l^z \rangle$ at $T/J=20$ are very
similar. 
At $T/J=0.2$ all autocorrelations show much more structure
 than at high $T$, but neither the $x$ autocorrelation (not shown) nor
 the $z$ autocorrelation (Fig.\ref{Fig18}, inset) show strong
 variations of the decay as $\Delta$ changes.

We now discuss 
the ``spatial variance'' $\sigma_{\mu}^2(t)$ as defined in 
(\ref{II.19}). For $\Delta=0$, $\sigma_{z}^2(t)$ is given by
(\ref{II.24}), and that
is precisely what the numerical
results show, apart from finite-size effects. 
For given $N$ and $T$, Re$ \sigma_z^2(t)/t^2$ and  Im$ 
\sigma_z^2(t)/t$ are constant up to some characteristic time 
$t_c$ which is approximately proportional to $N$ and grows 
slightly with $T$; at high $T$ we have $t_c \approx N/4J$. For 
$t>t_c$ finite-size effects start to build up. Assuming that 
$t_c$ is the time it takes the fastest excitations to travel the 
distance $N/2$, one obtains a velocity $v=2J$ for these 
excitations. 
This is  the maximum velocity of the 
Jordan-Wigner fermions 
and also the spinon velocity (\ref{II.9a})for 
$\gamma= \pi/2$. The root-mean-square
 velocity  $[\mbox{Re} \sigma_z^2(t)/t^2]^{1/2}$
(compare (\ref{II.21}))
is smaller (by a factor $\sqrt2$, see (\ref{II.23})) than the 
maximum velocity at high temperatures. With decreasing $T$,  Re$ 
\sigma_z^2(t)/t^2$ becomes smaller, whereas 
Im$ \sigma_z^2(t)/t$ grows, as may be seen by combining
 (\ref{II.20}), and a
high-temperature expansion to obtain
\begin{mathletters}
\begin{equation}
\frac{J^2}{2} - \frac{\mbox{Re} \sigma_z^2(t)}{t^2} \propto T^{-2}
\label{III.1a},
\end{equation}
and
\begin{equation}
\frac{\mbox{Im} \sigma_z^2(t)}{t} \propto T^{-1}
\label{III.1b}.
\end{equation}
\end{mathletters}

At $\Delta=\cos(0.3\pi)$, Re$ \sigma_z^2(t)/t^2$ and  Im$ 
\sigma_z^2(t)/t$ are not constant, but decrease slightly 
from their $t=0$ values, until finite-size effects set in. This 
decrease becomes stronger at $\Delta=1$. The dependence of the 
$t=0$ values of Re$ \sigma_z^2(t)/t^2$ and  Im$ \sigma_z^2(t)/t$ 
on $T$ is roughly similar for all three $\Delta$ values considered
here (see (\ref{III.1a},\ref{III.1b})) 
However, $\lim_{t \rightarrow 0} {\mbox{Re} \sigma_z^2(t)}/{t^2}$ 
is a non-monotone function of $T$ at $\Delta \neq 0$, with a 
maximum at intermediate $T$; at high $T$ it becomes roughly $T$-
independent.

The spatial variance $\sigma_x^2(t)$ of the $x$ correlation shows 
qualitatively the same behavior for all $T$ and $\Delta$ values 
studied here. Re$ \sigma_x^2(t)/t^2$ and  Im$ \sigma_x^2(t)/t$ 
both decrease from their $t=0$ values, with the decay becoming 
more rapid as one goes from  $\gamma=0$ to $\gamma=\pi/2$. The 
$t=0$ values of these quantities 
themselves drop rather rapidly as $\gamma$ grows. For 
$\gamma=0.3\pi$ the $T$-dependence of   $\lim_{t \rightarrow 0} 
{\mbox{Re} \sigma_x^2(t)}/{t^2}$ is similar to the $\gamma=0$ case, starting
with a small value at low $T$ and saturating at a higher value for
high $T$, 
whereas at $\gamma=\pi/2$  $\lim_{t \rightarrow 0} 
{\mbox{Re} \sigma_x^2(t)}/{t^2}$ starts out with a small value at 
low $T$ and then rapidly drops to 0. 
To summarize, the data for $\sigma_{\mu}^2(t)$ do not provide evidence
for diffusive behavior within the time range accessible to our
study. The behavior of $\sigma_{\mu}^2(t)$ does not deviate
dramatically from the form (\ref{II.24}) valid for $\Delta=0$ at
arbitrary times and for arbitrary $\Delta$ at short times. 

We finally discuss the intermediate structure factors $I^x(q,t) $ and
$I^z(q,t) $ as defined in (\ref{I.2}).  In the $XX$ model we find at
$T/J=20$ a $q$-independent  decay of $I^x(q,t)$, very similar to
the $T=\infty$ Gaussian.  The
deviations from the Gaussian decay which become visible at lower $T$
were already discussed above in terms of the spin pair
correlations. Similarly, the evaluation of $I^z(q,t)$ for the $XX$
model yields no new insights as compared to the spin pair correlation
or the dynamic structure factor.

\begin{figure}[h] 
\centerline{ \epsfxsize=9cm \epsfbox{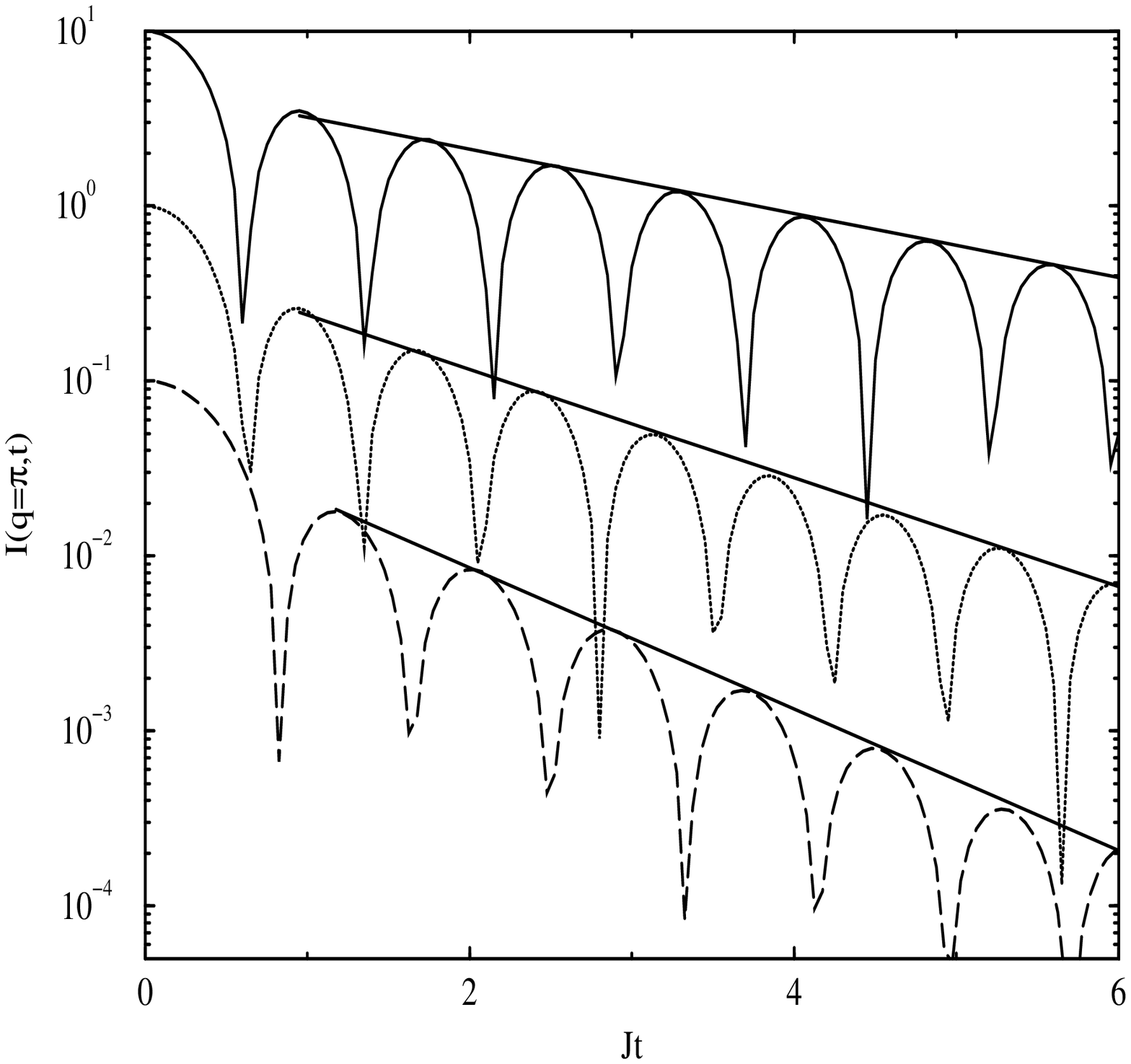} } 
\caption{The intermediate structure factor at $T=\infty$.
The solid curve represents $ |I^z(q=\pi, t, \Delta=\cos(0.3\pi))|$,
the dotted $|I(q=\pi, t, \Delta=1)|$ and the dashed 
$|I^x(q=\pi, t, \Delta=\cos(0.3\pi)|$.
For reasons of visibility the upper (lower) curve is multiplied by
a factor of 10 (0.1). The straight lines are drawn through the fourth
and fifth maxima.}
\label{Fig19} 
\end{figure}

For the general case of the $XXZ$ model, however, both  $I^x(\pi,t)$  and 
$I^z(\pi,t)$ show an exponential $t$ dependence for sufficiently large
$t$ and $T$ values \cite{note3}. 
This is shown in Fig.\ref{Fig19}, where we have
plotted the absolute values of $I^x(\pi,t)$ and $I^z(\pi,t)$ for
$T=\infty$ .

In the time range $Jt< 5 $ the $T/J=20$ curves coincide with the
$T=\infty$ curves. At $T/J=2$, the exponential decay 
is much less marked and for even smaller T it is no longer perceptible. 

\section{Summary}
\label{VI}

We have performed the first calculation of dynamic correlation
functions for the antiferromagnetic spin-1/2 $XXZ$ chain (\ref{I.1})
at arbitrary temperature by complete diagonalization of systems with
size $N \leq 16$ and anisotropies $\Delta=0, \cos(0.3\pi)$, and 1.
We have calculated the dynamic structure 
factors $S^{\mu}(q,\omega) \quad (\mu=x,z)$
(\ref{I.4}) and their Fourier transforms $\langle S^{\mu}_l(t)
S^{\mu}_{l+n} \rangle$ and $I^{\mu}(q,t)$ (\ref{I.2}).

Our study extends the ($T=0$, $N\leq 10$) work of
Refs.\onlinecite{MTBB,MTPB81} to larger systems and to $T > 0$. We
find that even at nonzero $T$ only a very small fraction of the many
possible transitions yield appreciable contributions to
$S^{\mu}(q,\omega)$ (see Sec. \ref{VA}). At all temperatures up to
$T=\infty$ the dynamic structure factor
$S^{z}(q,\omega)$ turns out to be of negligible size
for $|\omega|$ greater than the upper boundary of the known
\cite{MTBB,MTPB81} excitation continua.  This is demonstrated in
Figs.\ref{Fig6}-\ref{Fig8}; Table \ref{two} shows the the very small
fraction of spectral weight of $S^z(q,\omega)$
outside the two-spinon continuum at $T=\infty$.
There are analogous results for the transversal structure factor
$S^x(q,\omega)$ but this function becomes 
$q$-independent for $\Delta=0$.
 Similar conclusions about
the spectral range of $S^{\mu}(q,\omega)$ hold also for the frustrated
Heisenberg model (\ref{Hfrus}) (see Fig.\ref{Fig3}), and for the
Haldane-Shastry model (\ref{HHS}) (see Fig.\ref{Fig1}), where the
$T=\infty$ dynamic structure factor for $N=16$ 
vanishes strictly outside the two-spinon continuum.

The general shape of $S^{\mu}(q,\omega)$ at high temperatures
resembles that of the (analytic) $T=\infty$ results for
$\gamma=0.5\pi$, where $S^{x}(q,\omega)$ is a ($q$-independent)
Gaussian ``hill'' and $S^{z}(q,\omega)$ is defined on a
restricted frequency range (which grows with $q$) with inverse
square-root divergences at the boundaries which lead to a distinct
``trough'' shape. These basic patterns are encountered again at
$\gamma=0.3\pi$. At $\gamma=0, \quad S^{\mu}(q,\omega)$ is a
``hill'' for $q < \pi/2$ and a ``trough'' for $q > \pi/2$.

The low-temperature dynamic structure factors for small $\omega$ and
$q \approx \pi$ are described very well  by the field-theory result
(\ref{S1}) of Schulz \cite{S86} (see Fig.\ref{Fig9})
which  contains no upper frequency cutoffs and can not be used at 
high temperatures (see Fig.\ref{Fig10}).

Some results for space- and time-dependent correlations 
$\langle S_l^{\mu}(t) S_{l+n}^{\mu} \rangle$ are discussed in
Sec. \ref{III}. These correlations turn out to be $N$-independent over
a time interval growing with $N$. For $\gamma=0.5\pi$ the long-time
asymptotic behavior of $\langle S_l^{\mu}(t) S_{l+n}^{\mu} \rangle$ is
known (see Sec. \ref{II}), and under suitable circumstances that known
long-time behavior may be detected in an $N=16$ system
(Fig.\ref{Fig17}). However, reliable conclusions for the long-time
asymptotic behavior (including the absence or presence of spin
diffusion) at $\gamma < 0.5 \pi$ cannot be drawn from $N\leq 16$
data. For sufficiently high $T$ and large $q$, the intermediate
structure factor $I^{\mu}(q,t)$ for $\gamma=0$ and $0.3 \pi$ displays
an exponential time dependence (Fig.\ref{Fig19}).

\acknowledgments

We would like to express our gratitude to G.M\"uller 
for a careful reading of the manuscript, helpful comments and criticism. 
We acknowledge K.-H. M\"utter's critical reading of Sec. \ref{V}.
We thank M.B\"ohm and H.Leschke for making their original 
data for figures 1 and 3 in Ref.\onlinecite{BL92} ($T=\infty$ bounds on
$\langle S_l^z(t) S_{l+n}^z \rangle$ in the isotropic Heisenberg model)
available to us. U.L. acknowledges support by the DFG 
Graduiertenkolleg Festk\"orperspektroskopie at Dortmund.

\appendix

\section{Bethe Ansatz notation}
\label{A}

The eigenstates of the Hamiltonian (\ref{I.1}) expanded in the basis of 
states with fixed number $r$ of down spins at positions 
$ x_1,...,x_r$ have the coefficients
$$g(x_1,...,x_r)=$$
\begin{equation}
\sum \exp i\left( k_{i_1} x_1 + ... + k_{i_r} x_r
                             + \frac{1}{2} \sum_{k<l} \varphi_{{i_k}{i_l}}
                                  \right)
\label{A.1}.
\end{equation}
The sum extends over all permutations $i_1,...,i_r$ of the integers
$1,...,r$. The pseudomomenta $k_i$ and phases $\varphi_{j l}$ 
are solutions of the equations
\begin{equation}
 \cot \frac{\varphi_{j l}}{2} = \frac {\Delta \left(\cot \frac {k_j}{2} - 
                                        \cot \frac {k_l}{2} \right) }
                             {1+\Delta -(1-\Delta)\cot \frac{k_j}{2} 
                                                  \cot \frac{k_l}{2} }
\label{A.2}
\end{equation}
\begin{equation}
N k_m = 2 \pi \lambda_m + \sum_{l\neq m} \varphi_{m l} 
\label{A.3}.
\end{equation}
A state is determined by the set of integers $\lambda_1,...,\lambda_r$
with $0\leq \lambda_i < N$.
The corresponding energy eigenvalue is
\begin{equation}
 E=\frac{1}{2}\Delta(N-4r)+2\sum_{l=1}^{r} \cos k_l 
\label{A.4}
\end{equation}
and the momentum is
\begin{equation}
k= \frac{2\pi}{N} \sum_{m=1}^r \lambda_m.
\label{A.4a}
\end{equation}
The ground state in the subspace with spin $S$ (for $\Delta=1$) has 
\begin{equation}
 \lambda_1,...,\lambda_r  = S+1 , S+3 , \cdots , N-S-1 
\label{A.5}.
\end{equation}
The pseudomomenta $k_i$ are real if $\lambda_i-\lambda_j \geq 2$.
If in addition all $\lambda_i > 0$ 
the corresponding states are called class-C states \cite{G64}.
The remaining solutions are called bound states 
and have (with few exceptions) complex pseudomomenta forming 
(for large $N$) strings in the complex plane. 
The subspace $M_{AF}$ introduced in \cite{FaTa84}
to describe spinon excitations contains all states for which
$\frac {N}{2} - \nu_1 = finite$ for $N \rightarrow \infty$. 
$\nu_1$ is the number of strings of length one.
We have used Eqs.(\ref{A.2})-(\ref{A.4a}) to determine the complete
set of class C solutions for $\Delta =1$ and  $N \leq 16$ 
for the analysis described
in Sec. \ref{VA}.

%%%%%%%%%%%%%%%%%%%%%%%%%%%%%%%%%%%%%%%%%%%%%%%%%%%%%%%%%%%%%%%%%%%%
\begin{table}[b]
\begin{tabular}{|c|c|c|c|c|}
%\hline
&&&&\\
$ N,q  $     & $S_{11}$ & $S_{10}$ & $S_{01}$ & $S_{00}$ \\
&&&&\\
\hline
$16 , \pi$ & 
1.8990 $\cdot 10^{-1}$ & 
5.7144 $\cdot 10^{-2}$ & 
5.7144 $\cdot 10^{-2}$ &
6.9581 $\cdot 10^{-1}$ \\
%\hline
$12 , \pi$  &
3.5785 $\cdot 10^{-1}$  &
8.8507 $\cdot 10^{-2}$ &
8.8507 $\cdot 10^{-2}$ &
4.6514 $\cdot 10^{-1}$ \\
%\hline
$16, \frac{\pi}{2}$  &
1.2673 $\cdot 10^{-1}$   &
6.2628 $\cdot 10^{-2}$ &
6.2628 $\cdot 10^{-2}$ &
7.4802 $\cdot 10^{-1}$ \\
%\hline
 $12 , \frac{\pi}{2}$  & 
2.4249 $\cdot 10^{-1}$  &
1.0750 $\cdot 10^{-1}$  &
1.0750 $\cdot 10^{-1}$  &
5.4251 $\cdot 10^{-1}$   \\
%\hline
%\hline
\end{tabular}
\caption{Spectral characteristics of the structure factor of the
isotropic Heisenberg chain at $T=\infty$. $S_{\alpha \beta}$ is the
fraction of $S^z(q,\omega)$ corresponding to a transition from a class
$\alpha$ state to a class $\beta$ state. $\alpha=1$ denotes class C,
$\alpha=0$ denotes non-class C.}
\label{one}
\end{table}
\newpage
%%%%%%%%%%%%%%%%%%%%%%%%%%%%%%%%%%%%%%%%%%%%%%%%%%%%%%%%%%%%%%%%%%%%
%%%%%%%%%%%%%%%%%%%%%%%%%%%%%%%%%%%%%%%%%%%%%%%%%%%%%%%%%%%%%%%%%%%%
\begin{table}[b]
\begin{tabular}{|c|c|c|c|c|}
%\hline
&&&&\\
$q \cdot \frac{8}{\pi}$ & $\varepsilon_{2}^{XXX}(q)/J$ & $x_{out}^{XXX}(q)$ & 
$\varepsilon_{2}^{XXZ}(q)/J$ & $x_{out}^{XXZ}(q)$\\
&&&&\\
\hline%\hline
1 & 1.2258 & 4.1539 $\cdot 10^{-2}$ & 1.0522 & 4.5615 $\cdot 10^{-2}$\\
%\hline
2 & 2.4045 & 2.4019 $\cdot 10^{-2}$ & 2.0639 & 1.6524 $\cdot 10^{-2}$\\
%\hline
3 & 3.4908 & 9.5773 $\cdot 10^{-3}$ & 2.9964 & 8.0768 $\cdot 10^{-3}$\\
%\hline
4 & 4.4429 & 4.1566  $\cdot 10^{-3}$ & 3.8137 & 4.2199 $\cdot 10^{-3}$\\
%\hline
5 & 5.2243 & 1.7826 $\cdot 10^{-3}$ & 4.4844 & 2.0314 $\cdot 10^{-3}$\\
%\hline
6 & 5.8049 & 7.5846 $\cdot 10^{-4}$ & 4.9828 &8.9729  $\cdot 10^{-4}$\\
%\hline
7 & 6.1625 & 3.4219 $\cdot 10^{-4}$ & 5.2898 & 4.1120 $\cdot 10^{-4}$\\
%\hline
8 & 6.2832 & 2.4222 $\cdot 10^{-4}$ & 5.3934 & 2.7909 $\cdot 10^{-4}$\\
%\hline
\end{tabular}
\caption{Spectral characteristics of the structure factor of the
XXX- and XXZ-Heisenberg  chain (XXZ for $\gamma=0.3$) at $T=\infty$: 
upper boundary of the
two-spinon continuum and fraction of spectral weight outside the
continuum.}
\label{two}
\end{table}
\newpage
%%%%%%%%%%%%%%%%%%%%%%%%%%%%%%%%%%%%%%%%%%%%%%%%%%%%%%%%%%%%%%%%
\newpage
%%%%%%%%%%%%%%%%%%%%%%%%%%%%%%%%%%%%%%%%%%%%%%%%%%%%%%%%%%%%%%%%%%%%
\newpage
%%%%%%%%%%%%%%%%%%%%%%%%%%%%%%%%%%%%%%%%%%%%%%%%%%%%%%%%%%%%%%%%%%%%
\newpage
%%%%%%%%%%%%%%%%%%%%%%%%%%%%%%%%%%%%%%%%%%%%%%%%%%%%%%%%%%%%%%%%%%%%
\newpage
%%%%%%%%%%%%%%%%%%%%%%%%%%%%%%%%%%%%%%%%%%%%%%%%%%%%%%%%%%%%%%%%%%%%
\newpage
%%%%%%%%%%%%%%%%%%%%%%%%%%%%%%%%%%%%%%%%%%%%%%%%%%%%%%%%%%%%%%%%
\newpage
%%%%%%%%%%%%%%%%%%%%%%%%%%%%%%%%%%%%%%%%%%%%%%%%%%%%%%%%%%%%%%%%
\newpage
%%%%%%%%%%%%%%%%%%%%%%%%%%%%%%%%%%%%%%%%%%%%%%%%%%%%%%%%%%%%%%%%%%%%
\newpage
%%%%%%%%%%%%%%%%%%%%%%%%%%%%%%%%%%%%%%%%%%%%%%%%%%%%%%%%%%%%%%%%%%%%
\newpage
%%%%%%%%%%%%%%%%%%%%%%%%%%%%%%%%%%%%%%%%%%%%%%%%%%%%%%%%%%%%%%%%%%%%
\newpage
%%%%%%%%%%%%%%%%%%%%%%%%%%%%%%%%%%%%%%%%%%%%%%%%%%%%%%%%%%%%%%%%%%%%
\newpage
%%%%%%%%%%%%%%%%%%%%%%%%%%%%%%%%%%%%%%%%%%%%%%%%%%%%%%%%%%%%%%%%%%%%
\newpage
%%%%%%%%%%%%%%%%%%%%%%%%%%%%%%%%%%%%%%%%%%%%%%%%%%%%%%%%%%%%%%%%%%%%
\newpage
%%%%%%%%%%%%%%%%%%%%%%%%%%%%%%%%%%%%%%%%%%%%%%%%%%%%%%%%%%%%%%%%%%%%
\newpage
%%%%%%%%%%%%%%%%%%%%%%%%%%%%%%%%%%%%%%%%%%%%%%%%%%%%%%%%%%%%%%%%

%%%%%%%%%%%%%%%%%%%%%%%%%%%%%%%%%%%%%%%%%%%%%%%%%%%%%%%%%%%%%%%%%%%%
\newpage
%%%%%%%%%%%%%%%%%%%%%%%%%%%%%%%%%%%%%%%%%%%%%%%%%%%%%%%%%%%%%%%%%%%%
\newpage
%%%%%%%%%%%%%%%%%%%%%%%%%%%%%%%%%%%%%%%%%%%%%%%%%%%% 
\newpage
%%%%%%%%%%%%%%%%%%%%%%%%%%%%%%%%%%%%%%%%%%%%%%%%%%%%%%%%%%%%%%%%%%%%
\newpage
%%%%%%%%%%%%%%%%%%%%%%%%%%%%%%%%%%%%%%%%%%%%%%%%%%%%%
\newpage
%%%%%%%%%%%%%%%%%%%%%%%%%%%%%%%%%%%%%%%%%%%%%%%%%%%%%%%%%%%%%%%%%%%%
\newpage
%%%%%%%%%%%%%%%%%%%%%%%%%%%%%%%%%%%%%%%%%%%%%%%%%%%%%%%%%%%%%%%%%%%%
\newpage
%%%%%%%%%%%%%%%%%%%%%%%%%%%%%%%%%%%%%%%%%%%%%%%%%%%%%%%%%%%%%%%%%%%%
\newpage
%%%%%%%%%%%%%%%%%%%%%%%%%%%%%%%%%%%%%%%%%%%%%%%%%%%%%%%%%%%%%%%%

\begin{references}
\bibitem{18neu} At first sight there seems to be a contradiction 
between $\eta_z=2$ for $\gamma = \pi/2$ and the universal $t^{-
1}$ behavior of $\langle S_l^z(t) S_{l+n}^z \rangle$ for the $XX$ 
case discussed earlier. However, it must be kept in mind that the 
continuum field theory of Refs.\onlinecite{LP75} and  
\onlinecite{F78} is unable to account for lattice effects such as 
the existence of a high-frequency cutoff in the particle-hole 
spectrum of noninteracting fermions which causes the $t^{-1}$ 
contribution for the $XX$ case.
\bibitem{constfix} The constant $A_{\mu}(\gamma)$ in (\ref{S1}) was
fixed so that satisfactory agreement to the numerical data was
obtained for all $T \leq 2.7 J$ and both $q=\pi$ and $7 \pi / 8$.
\bibitem{note1} The relative deviation between the $N=16$ and
$N=\infty$ results is less than $10^{-5}$ for $Jt\leq 4.5$; it grows
to about $1.5 \cdot 10^{-3}$ for $Jt=5$.
\bibitem{note3} It should be remarked that for $Jt < 5$, the N=14
curves coincide with N=16, so in the time range $Jt < 5$ we can be
sure that our results represent the thermodynamic limit.
\end{references}
\end{document}